\documentclass[prc,twocolumn,letterpaper,nofootinbib,tightenlines,superscriptaddress,showpacs,preprintnumbers]{revtex4-1}

\usepackage{amsmath,amssymb,amsfonts}

\usepackage[svgnames]{xcolor}
\usepackage[english]{babel}
\usepackage{blindtext}
\usepackage{microtype}
\usepackage{tikz}
\usepackage{dcolumn}
\usepackage{multirow}
\usepackage{slashed}

\usepackage[caption=false]{subfig}

\usepackage[colorlinks=true]{hyperref}
\usepackage{ifpdf}
\usepackage{graphicx}
\graphicspath{{.}{figures/}}

\def\hs{\hspace}

\def\no{\nonumber}

\def\lf{\left}
\def\rg{\right}

\font\bb=bbmss10 scaled 1200
\def\ident{\mbox{\bb 1}}

\begin{document}

\title{Dressed Quark Mass Dependence of Pion and Kaon Form Factors}

\author{Y.~Ninomiya}
\email[Corresponding author:~]{3bsnm017@mail.tokai-u.jp}
\author{W.~Bentz}
\affiliation{Department of Physics, School of Science, Tokai University,
             4-1-1 Kitakaname, Hiratsuka-shi, Kanagawa 259-1292, Japan}
\author{I.~C. Clo\"et}
\affiliation{Physics Division, Argonne National Laboratory, Argonne, Illinois 60439, USA}

\begin{abstract}
The structure of hadrons is described well by the Nambu--Jona-Lasinio (NJL) model, which is a chiral effective 
quark theory of QCD. 
In this work we explore the electromagnetic structure of the pion and kaon using the 
three-flavor NJL model in the proper-time regularization scheme, including effects of the pion cloud 
at the quark level. 
In the 
calculation there is only one free parameter, which we take as the dressed light quark ($u$ and $d$) mass. 
In the regime where the dressed light quark mass is approximately $0.25\,$GeV, we find that the calculated 
values of the kaon decay constant, current quark masses, and quark condensates are consistent with experiment
and QCD based analyses. We also investigate 
the dressed light quark mass dependence of the pion and kaon electromagnetic form factors, where comparison 
with empirical data and QCD predictions also favors a dressed light quark mass near $0.25\,$GeV.
\end{abstract}


\maketitle

\section{Introduction}
Since the 1960s there have been substantial efforts, both experimental and theoretical, to unravel 
the quark structure of hadrons. The electromagnetic form factors of the various hadrons have played a crucial 
role in this process, as they reflect their internal quark (and gluon) 
structure~\cite{Thomas:2001kw,Arrington:2006zm,Cloet:2008re}. The form factors
of the pion and the kaon are of particular interest, because these mesons are associated with the Goldstone 
modes of dynamical chiral symmetry breaking~\cite{lee1972chiral} and play important roles in the description of 
the nuclear force~\cite{Machleidt:2011zz}. The pion form factor has been measured
in the region of low to medium momentum transfer~\cite{Amendolia:1986wj,Huber:2008id} and future
measures at higher momentum are planned~\cite{Huber:2006pac30}.
The kaon form factor, on the other hand, is poorly known experimentally, except in the region of
very low momentum transfer~\cite{Amendolia:1986ui}.
On the theoretical side, QCD based studies of the pion and kaon form factors have
been carried out, e.g., in the framework of perturbative QCD~\cite{Farrar:1979aw,Lepage:1980fj}, the Dyson-Schwinger equations
\cite{Alkofer:1993gu,Maris:2000sk,Chang:2013pq} and the Nambu-Jona-Lasinio (NJL) model~\cite{Blin:1987hw,Bernard:1988bx}.

The main purpose of this paper is to study the dressed (or equivalently constituent) quark mass dependence of the 
pion and kaon electromagnetic form factors,
including effects from the virtual pion cloud around the dressed quarks and from vector mesons, using the
three-flavor NJL model with four-fermion interactions.
The NJL model is a powerful chiral effective quark theory of QCD~\cite{Vogl:1991qt,Hatsuda:1994pi}, 
with numerous successes in the study
of meson~\cite{Klevansky:1992qe,Vogl:1991qt} and baryon~\cite{Ishii:1993np,Ishii:1993rt,Ishii:1995bu} structure. 
In several recent studies~\cite{Bentz:2001vc,Cloet:2005pp,Cloet:2007em,Roberts:2011cf,Segovia:2013uga} 
it has been demonstrated that one important aspect of quark confinement, namely, the absence
of thresholds for the decay of hadrons into free quarks, can
be implemented via a judicious choice for the regularization prescription.
Following these lines, we will use the proper-time 
scheme~\cite{Ebert:1996vx,Hellstern:1997nv,Bentz:2001vc} in this study. 
In our calculations there is only one free parameter, 
which we take as the dressed light quark ($u$ and $d$) mass $M$.\footnote{We assume isospin symmetry 
and denote $M_u=M_d=M$ for the dressed $u$ and $d$ quark masses, 
and $m_u = m_d = m$ for the associated current quark masses.}
The constituent quark model suggests dressed quark masses in the range $0.3\,$--$\,0.4\,$GeV, and it is often fixed at
$0.4\,$GeV in NJL model calculations of form factors~\cite{Horikawa:2005dh,Cloet:2014rja} and 
structure functions~\cite{Cloet:2005pp}. 
However, an important goal of our present study is to show that results for the current 
quarks masses, quark condensates, the kaon leptonic decay constant, as well as the pion and kaon charge radii and form factors,
can be improved by using a smaller dressed light quark mass of $M \sim 0.25\,$GeV. 
The dressed quark mass dependence of these observables is therefore investigated.
                                      
Recent experimental analyses of the current quark masses and pseudoscalar meson leptonic decay constants have 
found: $m_s/m = 27.5 \pm 1.0$~\cite{Beringer:1900zz} and $f_K/f_{\pi} = 1.197 \pm 0.002 \pm 0.006 \pm 0.001$
~\cite{Beringer:1900zz,Rosner:2013ica}. The lattice QCD calculations of Refs.~\cite{Durr:2010vn,Carrasco:2014cwa}, 
which are extrapolated to the continuum limit, find $m_s/m = 27.53 \pm 0.20 \pm 0.08$ and for the 
pseudoscalar decay constants Refs.~\cite{Dowdall:2013rya,Bazavov:2013vwa} obtain $f_K/f_\pi = 1.1916 \pm 0.0021$. 
Both these results are in excellent agreement with experiment. 
Concerning quark condensates, a recent lattice QCD analysis \cite{McNeile:2012xh} found  
$\left< \bar{s} s \right> / \left< \bar{\ell} \ell \right> = 1.08 \pm 0.16$ for the
ratio of strange to light ($\ell = u,d$) nonperturbative (physical) quark condensates.
As we shall see, our results for those three ratios $m_s/m$, $f_K/f_{\pi}$ and
$\langle \bar{s} s \rangle / \langle \bar{\ell} \ell \rangle$, together with the pion charge radius, are in 
excellent agreement with the empirical and QCD based results if the mass of the 
dressed light quark is approximately $M \sim 0.25\,$GeV. 
We emphasize that, because our model is free of unphysical decay thresholds, there are no problems 
in obtaining hadron masses which are greater than the sum of their dressed 
quark masses, which is important for the extension of these studies to, e.g., the $\rho$ meson and the nucleon.
The main point which we wish to make in this paper is to show that the results for the pion and kaon form factors, 
as well as the other physical quantities mentioned above, can be much improved by using a rather small value of 
the dressed light quark mass. 
Corrections from the pion cloud and vector mesons to the quark-photon vertex 
are important to attain this good overall picture.         
In order to keep the simplicity of the NJL model description, however, we do not intend to present 
a full study of meson loops in this work. We will explicitly consider the 
meson loop corrections only for those processes which involve an external virtual photon probe, 
that is, the pion cloud corrections (see Fig.~\ref{PC-figure}) and the corrections from 
$\rho$ and $\omega$ mesons (see Fig.~\ref{VMD}) to the quark-photon vertex. For the other processes,         
like those shown in Figs.~\ref{Gap eq}, \ref{ladder}, \ref{bubble graph}, and \ref{decay constant} only that part of the pion 
loop effects which can be incorporated 
into the mass and wave function renormalization of the dressed quarks will be considered, which is 
sufficient to uphold various important low energy theorems (see App. C). 
This is essentially the same kind of approximation which has been implicitly used in numerous works on 
meson cloud and meson exchange current effects in hadronic~\cite{Suzuki:1997wv,Szczurek:1996tp,Cloet:2014rja} and nuclear~\cite{Towner:1987zz,Arima:1988xa} physics. 
Nevertheless, a more complete study 
of meson loops, including their effects also on processes without an external probe~\cite{Nikolov:1996jj,Dmitrasinovic:1995cb}, should be  
an important goal for future studies.    


The outline of this paper is as follows: In Sect.~\ref{sec:njl} we introduce the model and provide expressions that 
give the current quark masses, the masses of pion and kaon, and their leptonic decay constants.  
In Sect.~\ref{sec:meson} we calculate the pion and kaon form factors, and Sect.~\ref{sec:results}
presents these results. A summary is given in Sect.~\ref{sec:summary}.
\section{The NJL model\label{sec:njl}} 
The NJL model~\cite{Nambu:1961tp,Nambu:1961fr} is a successful chiral effective quark theory of QCD, that has been 
used to describe low to medium energy phenomena, such as dynamical chiral symmetry breaking 
and the associated dynamical quark mass generation. 
In this section we briefly explain the 
three-flavor NJL model with four-fermion interactions, together with the proper-time 
regularization scheme which avoids unphysical decay thresholds. 
We also illustrate the relation
between the dressed and current quark masses, and discuss mesons as relativistic bound states 
of a dressed quark and anti-quark.

\subsection{NJL Lagrangian and the Gap Equation}
The three-flavor NJL model Lagrangian, with four-fermion interactions, reads   
\begin{align}
\hspace*{-1mm}\mathcal{L}_{NJL} &= \bar{\psi}(i\slashed{\partial} - \hat{m})\psi 
+ G_{\pi}\left[(\bar{\psi}\,\lambda_{a}\,\psi)^{2} - (\bar{\psi}\,\lambda_{a}\,\gamma_{5}\,\psi)^{2}\,\right] \no \\
&\hs{12mm}
- G_v\left[(\bar{\psi}\,\lambda_{a}\,\gamma^\mu\,\psi)^{2} + (\bar{\psi}\,\lambda_{a}\,\lambda_{a} \gamma_{5}\,\psi)^{2}\right],
\label{NJL lagrangian}
\end{align}
where the quark field $\psi$ has the flavor components $\psi = (u, d, s)$ and $\hat{m}$ denotes the
current quark mass matrix $\hat{m}={\rm diag} (m, m, m_s)$. A sum over $a = 0,\ldots,8$ is implied in 
Eq.~\eqref{NJL lagrangian}, where $\lambda_1,\dots,\lambda_8$ are the Gell-Mann matrices in flavor space
and $\lambda_0 \equiv \sqrt{\frac{2}{3}}\,\ident$.
To explicitly break the global $U_A(1)$ symmetry of Eq.~\eqref{NJL lagrangian} and describe, for example, 
also the $\eta$ and $\eta'$ mesons, a six-fermion (determinant) interaction~\cite{'tHooft:1976fv} 
is often included in Eq.~\eqref{NJL lagrangian}. However, because this term
will not directly affect our main results on pion and kaon properties, we do not include it here for simplicity.\footnote{As pointed out in Ref.~\cite{Osipov:2006ns}, in order to avoid an unstable vacuum, the inclusion of the
six-fermion interaction makes it necessary to also include an eight-fermion interaction.
In order to retain the simplicity of the model, we do not include these interactions in this work.}
The four-fermion interaction term proportional to the coupling constant $G_{\pi}$ in Eq.~\eqref{NJL lagrangian} 
describes the direct terms of the $\bar{q}q$ interaction in the scalar and pseudoscalar meson channels. This term is
responsible for, \textit{inter alia}, the dynamical breaking of chiral symmetry and consequentially the generation 
of dressed quark masses. 
The term proportional to $G_v$ in Eq.~\eqref{NJL lagrangian} describes the direct piece of the $\bar{q}q$ interaction 
in the vector and axialvector meson channels.\footnote{In principle the flavor singlet and octet pieces of the
$G_v$ term in Eq.~\eqref{NJL lagrangian} can appear in the NJL interaction Lagrangian with 
separate coupling constants, as they are individually chirally symmetric. Our choice of identical
coupling constants avoids flavor mixing, giving the $\omega$ meson as $(u \bar{u} + d \bar{d})$ and the
$\phi$ meson as $s \bar{s}$.}
The NJL model does not \textit{a priori} contain quark confinement.  
However, one important aspect of quark confinement 
can be incorporated into the NJL model by introducing an infrared cut-off in the proper-time regularization 
scheme~\cite{Ebert:1996vx,Hellstern:1997nv,Buck:1992wz,Bentz:2001vc}. 
This additional cut-off eliminates unphysical thresholds for the decay of hadrons into quarks, and
at the same time respects all symmetry 
constraints. (Details of this regularization method are discussed further in App.~\ref{app:pt}.)


\begin{figure}[tbp]
\centering\includegraphics[width=0.5\columnwidth,clip=true,angle=0]{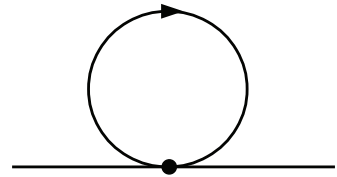}
\caption{Quark self-energy in the mean-field approximation. The solid line represents a dressed quark propagator.}
\label{Gap eq}
\end{figure}

In the mean-field approximation 
the dressed quark masses ($M$ and $M_s$) are given by the quark self-energy 
illustrated in Fig.~\ref{Gap eq}. Because the relevant interaction term in the NJL Lagrangian~(Eq.~\eqref{NJL lagrangian})
is given by  
\begin{align}
G_{\pi}\sum_{a=0,3,8} \left(\bar{\psi} \lambda_a \psi\right)^2 = 2 G_{\pi} \left[
\left(\bar{u} u\right)^2 + \left(\bar{d} d\right)^2 + \left(\bar{s} s\right)^2 \right],
\end{align}
the gap equations decouple in flavor space and take the familiar forms 
\begin{align}
M_q &= m_q - 4\,G_{\pi} \left< \bar{q}q \right> \no\\
    &= m_q + 48i\,G_{\pi}\,M_q \int \frac{d^{4}k}{(2\pi)^{4}}\ \frac{1}{k^2 - M_q^2 + i \epsilon}\,,
\label{constituent quark mass}
\end{align}
where $q = u,\,d,\,s$, and $\left< \bar{q}q \right>$ is the quark condensate.   
Using a Wick rotation and introducing the proper-time regularization gives
\begin{align}
\frac{m_q}{M_q} &= 1 - \frac{3\,G_{\pi}}{\pi^{2}}\int^{1/\Lambda_{IR}^{2}}_{1/\Lambda_{UV}^{2}} d \tau \
\frac{e^{-\tau M_q^2}}{\tau^2}.
\end{align}
The dressed quark propagators for the light and strange quarks are therefore given respectively by
\begin{align}
S_\ell(p)   &= \frac{\slashed{p}+M} {\lf[p^2 - M^2 + i\varepsilon\rg]},
\label{propagator}
 \\
S_s(p) &=\frac{\slashed{p}+M_s}{ \lf[p^2 - M_s^2 + i\varepsilon\rg]},
\end{align}
and in flavor space the quark propagator has the form
\begin{align}
S(p) = \text{diag}\lf[S_\ell(p),\,S_\ell(p),\,S_s(p)\rg].
\label{eq:flavour_space_quark_propagator}
\end{align}

\subsection{Mesons and their couplings to quarks \label{mesons couplings}}
\begin{figure}[tbp]
\centering\includegraphics[width=\columnwidth,clip=true,angle=0]{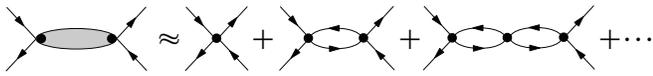}
\caption{Random phase approximation for the quark-antiquark $T$-matrix.}
\label{ladder}
\end{figure}
\begin{figure}[tbp]
\centering\includegraphics[width=\columnwidth,clip=true,angle=0]{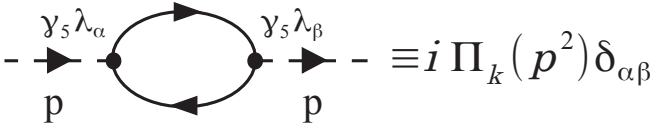}
\caption{The pseudoscalar meson bubble diagram, $\Pi_k(p^2)$, where $k=\pi,\,K$. The dashed line represents a pion 
($\alpha, \beta = 1,2,3$) or a kaon ($\alpha, \beta = 4,5,6,7$).}
\label{bubble graph}
\end{figure}

The pion and kaon $T$-matrices are obtained by considering quark--antiquark scattering in the pseudoscalar 
channel using the random phase approximation (RPA), which is equivalent to the ladder approximation, and 
is illustrated in Fig.~\ref{ladder}. Summing the bubble diagrams in Fig.~\ref{ladder} gives
\begin{align}
T_k = \gamma_{5}\lambda_{\alpha} \ \frac{-2i\,G_{\pi}}{1 + 2\,G_{\pi}\,\Pi_k(p^{2})} \ \gamma_{5}\lambda_{\alpha},
\label{T-matrix pion}
\end{align}
where $k = \pi, K$; the sum over $\alpha$ takes the values $\alpha = 1,2,3$ for the pion ($k=\pi$) and 
$\alpha = 4,5,6,7$ for the kaon ($k=K$). The matrices $\gamma_5 \lambda_{\alpha}$ act on the external quarks,
and $\Pi_k(p^2)$ is the bubble diagram (polarization propagator) in the pion and kaon channels, 
illustrated in Fig.~\ref{bubble graph}. These bubble diagrams take the form
\begin{align}
\Pi_{\pi}(p^2)\,\delta_{\alpha\beta}
&= i\int\frac{d^4 k}{(2\pi)^4}\ {\rm Tr}\lf[\gamma_{5}\,\lambda_{\alpha}\,S_\ell(p+k)\,\gamma_{5}\,\lambda_{\beta}\,S_\ell(k)\rg], \\
\Pi_{K}(p^2)\,\delta_{\alpha\beta}
&= i\int\frac{d^4 k}{(2\pi)^4}\,{\rm Tr} \lf[\gamma_{5}\,\lambda_{\alpha}\,S_\ell(p+k)\,\gamma_{5}\,\lambda_{\beta}\,S_s(k)\rg], 
\end{align}
where for the pion $\alpha, \beta=1,\,2,\,3$ and for the kaon $\alpha, \beta = 4,\,5,\,6,\,7$. The 
trace is taken in Dirac, flavor and color space. Explicit forms for these bubble diagrams, in the
proper-time regularization scheme, are given in App.~\ref{app:bubble}.

The pion and kaon masses, $m_k$, are defined by the pole in the corresponding $T$-matrix, therefore
the pole conditions take the form 
\begin{align}
1 + 2\,G_{\pi}\,\Pi_k\!\lf(p^{2}=m_k^2\rg)=0, \quad \text{where} \quad k = \pi,\,K.
\label{pipole}
\end{align}
Near a bound state pole the $T$-matrix behaves as
\begin{align}
T_M \sim \gamma_{5}\lambda_{\alpha} \ \frac{ig_k^{2}}{p^{2}-m_k^2 + i \epsilon} \ \gamma_{5}\lambda_{\alpha},
\label{T-matrix2}
\end{align}
%
where $g_k$ is identified as the quark--meson coupling constant. 
To derive expressions for $g_k$ we expand
Eq.~\eqref{T-matrix pion} about the pole at $p^2 = m_k^2$. Using
\begin{align}
\Pi_k(p^{2}) &= \Pi_k(m_k^{2}) + \lf.\frac{\partial \Pi_k(p^{2})}{\partial\, p^{2}}\rg|_{p^{2}=m_k^2}(p^{2}-m_k^{2}) + \ldots,
\end{align}
gives
\begin{align}
g_k^2 = -\lf[\lf.\frac{\partial \Pi_k(p^{2})}{\partial p^{2}}\rg|_{p^2=m_k^2}\rg]^{-1}.
\label{kaon quark constant}
\end{align}
From the pole behavior in Eq.~\eqref{T-matrix2} we see that the quark-antiquark interactions are mediated by 
pseudoscalar particles. Hence, we can interpret $m_k$ as the meson mass and $g_k$ as the coupling constant 
of the meson to the quarks.
We will use the pole approximation for the $T$-matrix, expressed by Eq.~\eqref{T-matrix2}, 
throughout this work in order to keep meson loop integrals tractable analytically.
The simple ladder approximation used here leads to pseudoscalar ($\gamma_5$) couplings of the
pion or kaon to the quarks. It is well known~\cite{Bernard:1993rz} that also a mixing
between the pseudoscalar and pseudovector interaction terms of the Lagrangian~\eqref{NJL lagrangian} 
can contribute
to the $T$-matrix in the pseudoscalar channel, which leads to a pseudovector contribution 
($\slashed{p} \gamma_5$) to the meson--quark coupling. Because this mixing is
physically associated with the contribution of a heavy meson (the $a_1$ meson for the light flavor case) 
in the intermediate states, we neglect it here so as to keep the simplicity of the model description
\footnote{For the case of the T-matrix in the pion channel, those mixing contributions
are proportional to $p^2$ and therefore expected to be small near the pion pole.
The mixing contributions to the pion form factor, however, may become
important for high values of $Q^2$.}.


\subsection{Meson decay constants}
\begin{figure}[bp]
\centering\includegraphics[width=\columnwidth,clip=true,angle=0]{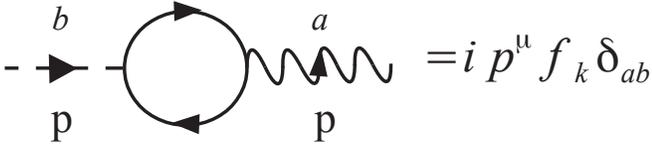}
\caption{Diagram representing the pseudoscalar meson decay constant. The dashed 
line represents a pseudoscalar meson and the wavy line an external axialvector field.}
\label{decay constant}
\end{figure}
The pion and kaon leptonic decay constants can be determined from the meson 
to hadronic vacuum matrix element, $\left<0|j_{a}^{\mu}(0)|k_{b}(p)\right>$ ($k=\pi,\,K$), 
where $j_a^{\mu}(x)$ is the weak axialvector current operator for flavor quantum number $a$.
This matrix element is illustrated diagrammatically in Fig.~\ref{decay constant}, 
and therefore the pion and kaon leptonic decay constants, $f_k$, are defined by
\begin{align}
\left<0|j_{a}^{\mu}(0)|k_{b}(p)\right> \equiv i\,p^{\mu}\,f_k\,\delta_{ab}.
\end{align}
The diagram in Fig.~\ref{decay constant} gives, for the kaon
\begin{align}
i\,p^{\mu}\,f_{K}\,\delta_{ab} &= -\,g_K \int\frac{d^4 k}{(2\pi)^4} \no \\
&\hs{2mm}\times
\mathrm{Tr}\lf[\tfrac{1}{2}\gamma^\mu\,\gamma_5\,\lambda_a\,S(k+p)\gamma_5\,\lambda_b\,S(k)\rg],
\end{align}
where $a, b = 4,5,6,7$; the trace is over Dirac, color and flavor space; and the 
quark propagator is given by Eq.~\eqref{eq:flavour_space_quark_propagator}.
Therefore
\begin{align}
f_{K} &= -12i\,g_{K} \int\frac{d^4 k}{(2\pi)^4}\frac{M_{s}+\frac{p\cdot k}{p^2}(M_{s}-M)}{[(p+k)^2-M_s^2][k^2-M^2]}.
\label{eq:kaon decay}
\end{align}
Introducing Feynman parameters gives
\begin{align}
f_{K}  &= -12i\,g_K\int^{1}_{0}dx\int\frac{d^4 k}{(2\pi)^4} \no \\
&\hs{1mm}\times \frac{M_{s}-x(M_{s}-M)}{[k^2 + x(1-x)\,m_K^2 - M_{s}^2 + x(M_{s}^2-M^2)]^2}.
\label{eq:kaon decay 1}
\end{align}
By Wick rotating and introducing the proper-time regularization scheme we find
\begin{align}
f_{K} &=\frac{3\,g_{K}}{4\pi^2}\int^{1}_{0}dx \int^{1/\Lambda_{IR}^{2}}_{1/\Lambda_{UV}^{2}} d\tau\ \frac{1}{\tau}\ \lf[M_s + x\,(M-M_s)\rg] \no\\
&\hs{22mm}\times e^{-\tau [M_{s}^2 - x(M_{s}^2-M^2) - x(1-x)m_K^2]}. 
\label{eq:fK} 
\end{align}
The result for $f_\pi$ is obtained from Eq.~\eqref{eq:fK} via the substitutions: $M_s \to M$, $g_K \to g_\pi$
and $m_K \to m_\pi$.

\section{Pion and Kaon form factors\label{sec:meson}}
The electromagnetic current, $j^{\mu}(p',p)$, of a hadron is defined by
\begin{multline}
\int d^{4}z\,e^{-iqz}\left<\vec{p}\,'\left|\bar{\psi}(z)\,\tfrac{1}{2}\!\left(\lambda_{3}+\tfrac{1}{\sqrt{3}}\lambda_{8}\right)\gamma^{\mu}\,\psi(z)\right|\vec{p}\right>\\
\equiv \sqrt{4\,E_{p}\,E_{p'}}\ (2\pi)^4\,\delta^{(4)}(p'-p-q)\,j^{\mu}(p',p),
\end{multline}
where $E_p = \sqrt{\vec{p}^{\,2} + m_k^2}$, $q = p'- p$ and the normalization of state vectors is
\begin{align}
\left<\vec{p}\,'|\vec{p}\,\right> = 2\,(2\pi)^{3}\,E_{p}\,\delta^{(3)}(\vec{p}\,'-\vec{p}).
\end{align}
For the case of a pseudoscalar meson, the electromagnetic current is parameterized by a single
form factor and takes the form
\begin{align}
\sqrt{4\,E_{p}\,E_{p'}}\ j_k^{\mu}(p',p) \equiv \lf(p'^{\mu} + p^{\mu}\rg) F_k(Q^2),
\end{align}
where $Q^2\equiv -q^2$. 

In the NJL model considered here the pion and kaon electromagnetic current is given
by the two diagrams of Fig.~\ref{vertex correction}; and in this section we determine 
the pion and kaon form factors at three levels of sophistication.
Firstly the pseudoscalar form factors are obtained by treating the dressed quarks like
point (bare) particles; in the second case a pion loop on the dressed quarks is included; and finally at the third
level of sophistication we also include vector meson contributions to the quark-photon vertex.

\subsection{Pion and Kaon form factors: bare quarks}                              
\begin{figure}[tbp]
\begin{center}
\includegraphics[width=\columnwidth,clip=true,angle=0]{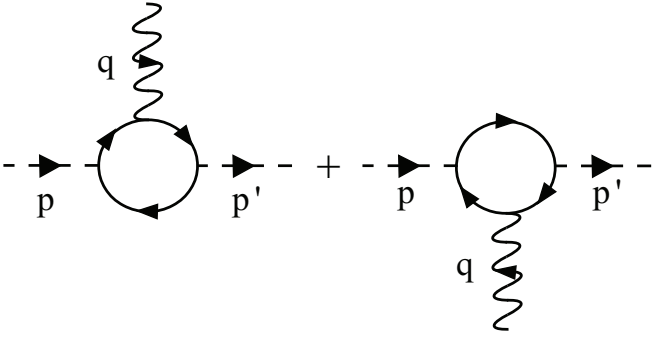}
\caption{Feynman diagrams for the meson electromagnetic current.}
\label{vertex correction}
\end{center}
\end{figure}

The coupling of a photon to a point-like (bare) quark is given by
\begin{align}
\hs*{-0.7mm}\Lambda_q^{\mu,\text{(bare)}} = \frac{1}{2}\!\left(\lambda_{3} + \frac{1}{\sqrt{3}}\lambda_{8}\right)\gamma^{\mu}
= \begin{pmatrix} \frac{1}{6} + \frac{\tau_3}{2} & 0 \\ 0 & e_s \end{pmatrix}\gamma^{\mu},
\label{eq:barevertex}
\end{align}
where $\tau_3$ is a Pauli matrix and $e_s$ is the $s$ quark charge.
With the quark-photon vertex given by Eq.~\eqref{eq:barevertex} the electromagnetic current 
of the $\pi^+$, obtained from the diagrams in Fig.~\ref{vertex correction}, reads
\begin{align}
j^{\mu,\text{(bare)}}_{\pi}(p',p) = j^{\mu,\text{(bare)}}_{\pi,1}(p',p) + j^{\mu,\text{(bare)}}_{\pi,2}(p',p),
\label{eq:barepioncurrent}
\end{align}
where
\begin{align}
\label{eq:jpi1}
&j^{\mu,\text{(bare)}}_{\pi,1}(p',p)=\frac{i\,g_{\pi}^2}{\sqrt{4\,E_{p}\,E_{p'}}}\int\frac{d^4 k}{(2\pi)^4} \no \\ 
&\hs{2mm}
{\rm Tr}\lf[\gamma_{5}\,\tau_{-}\,S(k+p')\,\Lambda_q^{\mu,\text{(bare)}}\,S(k+p)\,\gamma_{5}\,\tau_{+}\,S(k)\rg], \\
\label{eq:jpi2}
&j^{\mu,\text{(bare)}}_{\pi,2}(p',p)=\frac{i\,g_{\pi}^2}{\sqrt{4\,E_{p}\,E_{p'}}}\int\frac{d^4 k}{(2\pi)^4} \no \\ 
&\hs{2mm}
{\rm Tr}\lf[\gamma_{5}\,\tau_{+}\,S(k-p)\,\Lambda_q^{\mu,\text{(bare)}}\,S(k-p')\,\gamma_{5}\,\tau_{-}\,S(k)\rg].
\end{align}
Note, the first term of the current corresponds to the left diagram in Fig.~\ref{vertex correction} 
and $j^{\mu,\text{(bare)}}_{\pi,2}$ the right diagram.
The flavor matrices from the Bethe-Salpeter vertices in Eqs.~\eqref{eq:jpi1}--\eqref{eq:jpi2}
are defined as $\tau_{\pm} \equiv \frac{1}{\sqrt{2}}(\lambda_{1} \pm i\lambda_{2})$ and
the quark propagator is given by Eq.~\eqref{eq:flavour_space_quark_propagator}.\footnote{For 
the $\pi^+$, with $M_u = M_d$, the two pieces of the current are related by 
$e_u^{-1}\,j^{\mu}_{\pi,1} = - e_d^{-1}\,j^{\mu}_{\pi,2}$ and could therefore 
be written as a single term.} 
The $K^+$ electromagnetic current reads
\begin{align}
j^{\mu,\text{(bare)}}_{K}(p',p) = j^{\mu,\text{(bare)}}_{K,1}(p',p) + j^{\mu,\text{(bare)}}_{K,2}(p',p),
\label{eq:barepioncurrent}
\end{align}
where $j^{\mu,\text{(bare)}}_{K,1}$ and $j^{\mu,\text{(bare)}}_{K,2}$ are obtained from Eqs.~\eqref{eq:jpi1} and
\eqref{eq:jpi2}, respectively, via the substitutions $g_\pi \to g_K$ and 
$\tau_{\pm} \to \lambda_{\pm} \equiv \frac{1}{\sqrt{2}}(\lambda_{4} \pm i\lambda_{5})$.

Taking the trace and introducing Feynman parameters, the (bare) pion and kaon form factors are given by
\begin{align}
\label{P-form factor}
&F_{\pi}^{\text{(bare)}}(Q^2) = 24i\,g_{\pi}^2\int\frac{d^4 k}{(2\pi)^4} \int^{1}_{0} dx \no \\
&\hs{8mm}
\times
\left[\frac{-x}{[k^2-\triangle_{1}]^2} + \frac{1}{2}\,m_{\pi}^2 \int^{x}_{-x} dy\ \frac{x}{[k^2-\triangle_{2}]^3} \right], 
\displaybreak
\\[0.3em]
\label{K-form factor}
&F_{K}^{\text{(bare)}}(Q^2) = 8i\,g_{K}^2\int\frac{d^4 k}{(2\pi)^4} \int^{1}_{0}dx \no \\
&\hs{6mm}
\times \left\{\frac{-2x}{[k^2-\triangle_{1}]^2}+\frac{-x}{[k^2-\triangle_{3}]^2}\right. \no \\
&\hs{15mm}
\left. + \int^{x}_{-x}dy \left[\frac{2N_{1}}{[k^2-\triangle_{4}]^3}+\frac{N_{2}}{[k^2-\triangle_{5}]^3}\right] \right\},
\end{align}
with
\begin{align}
\label{triangle1}
\triangle_{1} &= M^2 + x(1-x)\,Q^2, \\
\triangle_{2} &= M^2 - x(1-x)\,m_{\pi}^2 + \frac{1}{4}Q^2(x^2-y^2), \\
\triangle_{3} &= M_{s}^2 + x(1-x)\,Q^2, \\
\triangle_{4} &= x\,M^2 + (1-x)\lf(M_{s}^2 - x\,m_{K}^2\rg) + \frac{Q^2}{4}(x^2-y^2), \\
\label{triangle5}
\triangle_{5} &= x\,M_{s}^2 + (1-x)\lf(M^2 - x\,m_{K}^2\rg) + \frac{Q^2}{4}(x^2-y^2),
\end{align}
and
\begin{align}
N_{1} &= (1-x)\,M M_s - M_s^2 + \frac{x}{2}\lf(M^2 + M_s^2 + m_{K}^2\rg), \\
N_{2} &= (1-x)\,M M_s  - M^2 + \frac{x}{2}\lf(M^2 + M_s^2 + m_{K}^2\rg).
\end{align}
In the limit where $M=M_{s}$, and therefore $m_{\pi}=m_{K}$ and $g_\pi = g_K$, the pion 
and kaon from factors are identical.

\subsection{Pion and Kaon form factors: pion cloud}

\begin{figure}[tbp]
\begin{center}
\includegraphics[width=\columnwidth,clip=true,angle=0]{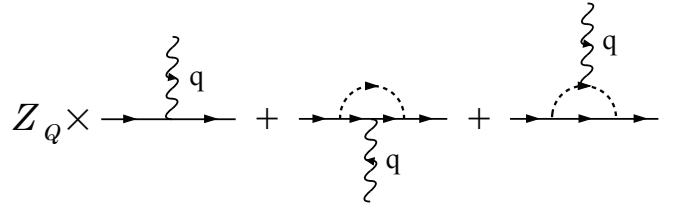}
\caption{Feynman diagrams for the quark electromagnetic current with a pion cloud.}
\label{PC-figure}
\end{center}
\end{figure}

In the previous subsection, we treated the coupling of the photon to the dressed quarks as point-like.
In general, however, the constituent quarks are dressed by a cloud of mesons. Because the pion is the 
lightest meson, effects of the pion cloud can contribute significantly to meson 
form factors for $Q^2 \lesssim 1\,$GeV$^2$~\cite{Cloet:2014rja}. Because of isospin conservation the $s$-quark cannot 
be dressed by the pion cloud and  therefore the pion cloud contribution to the pion form factor will
be about twice that for the kaon form factor. In this subsection we consider corrections to the 
quark-photon vertex from pion loops around a constituent quark, as illustrated in Fig.~\ref{PC-figure},
and determine their contribution to pion and kaon form factors.

As we mentioned already in Sect.~I, a full treatment of meson cloud effects is
very complicated and beyond the scope of this work. Here we will follow the
procedure explained in Ref.~\cite{Horikawa:2005dh,Mineo:1999eq}, which has been used implicitly in many previous
works and which we summarize in App.~\ref{app:renormalize}, to incorporate 
a part of pion cloud effects into a renormalization of the mass and wave function of the light
dressed quarks.
\footnote{
In App.~\ref{app:renormalize}, we also discuss the validity of low-energy theorems relevant in the present context, like the
Goldberger-Treiman (GT) relation~\cite{Goldberger:1958tr} or the Gell-Mann - Oakes - Renner (GOR) relation\cite{GellMann:1968rz}.}
Only for the quantity which is of most interest to our present work, 
namely the electromagnetic quark-photon vertex, the pion cloud effects are resolved and
explicitly treated.   

\begin{figure}[tbp]
\begin{center}
\includegraphics[width=\columnwidth,clip=true,angle=0]{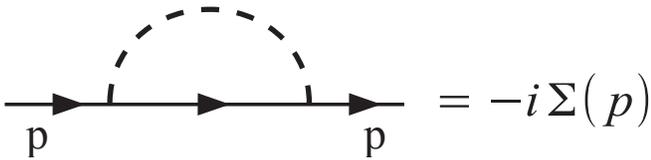}
\caption{Pion cloud self-energy diagram for the light quarks.}
\label{SelfEnergy}
\end{center}
\end{figure}

Including pion loop corrections modifies the flavor $SU(2)$ piece of Eq.~\eqref{eq:barevertex}, 
such that\footnote{Contributions
from a kaon cloud would modify each piece of Eq.~\eqref{eq:barevertex}.}
\begin{multline}
\left(\frac{1}{6}+\frac{\tau_{3}}{2}\right)\gamma^{\mu} 
\longrightarrow \
Z_{Q}\left(\frac{1}{6}+\frac{\tau_{3}}{2}\right)\gamma^{\mu} \\
+ \frac{1}{2}\,(1-\tau_3)\,\Lambda_{Q}^{\mu}(p',p) + \tau_3\,\Lambda_{\pi}^{\mu}(p',p),
\label{eq:vertexpion}
\end{multline}
where each term is associated with the corresponding diagram in Fig.~\ref{PC-figure}. The quark wave 
function renormalization, $Z_{Q}$, is essential for charge conservation
and is interpreted as 
the probability of striking a dressed quark without its pion cloud. It is given by
(see App. \ref{app:renormalize}) 
\begin{align}
Z_{Q} = 1 + \lf.\frac{\partial \Sigma (p)}{\partial \slashed{p}}\rg|_{\slashed{p}=M},
\label{renormalization constant}
\end{align}
where $\Sigma(p)$ is the light quark self-energy arising from the pion cloud, illustrated 
in Fig.~\ref{SelfEnergy}. This self-energy reads
\begin{align}
\Sigma(p) = 3i\,g_{\pi}^2\int \frac{d^4k}{(2\pi)^4} \ iD_{\pi}(p-k)\,\gamma_{5}\,iS_\ell(k)\,\gamma_{5},
\end{align}
where $D_{\pi}(p)$ denotes the pion propagator given by
\begin{align}
D_{\pi}(p) = \frac{1}{p^2 - m_{\pi}^2 + i\varepsilon}.
\end{align}

The vertex functions of Eq.~\eqref{eq:vertexpion} take the form
\begin{align}
\label{Q-formfactor-temporary}
\Lambda_{Q}^{\mu}(p',p) &= g_{\pi}^2 \int \frac{d^4k}{(2\pi)^4}  \no \\
&\hs{5mm}\gamma_{5}\,iS_\ell(p'-k)\,\gamma^{\mu}\,iS_\ell(p-k)\,\gamma_{5}\,iD_{\pi}(k), \\
\label{Pion-photon-C}
\Lambda_{\pi}^{\mu}(p',p) &= 2\,g_{\pi}^2\,\lf(p'^\mu+p^\mu\rg)F_{\pi}^{\text{(bare)}}(q^2)\, \int \frac{d^4k}{(2\pi)^4} \no \\
&\hs{9mm}iD_{\pi}(p'-k)\,iD_{\pi}(p-k)\,\gamma_{5}\,iS(k)\,\gamma_{5},
\end{align}
where $p'$ and $p$ are the external momenta of the quarks. 
The off-shell vertex functions of Eqs.~\eqref{Q-formfactor-temporary}--\eqref{Pion-photon-C}
are approximated by their on-shell form in our calculation of the meson form factors.
The vertex functions in Eq.~\eqref{eq:vertexpion} can therefore be expressed in the form
\begin{align}
\label{FQ_define}
\Lambda_{Q}^{\mu}(p',p) &= \gamma^{\mu}F^{(q)}_{1Q}(Q^2) + \frac{i\sigma^{\mu\nu}q_{\nu}}{2M}F_{2Q}^{(q)}(Q^2), \\
\label{FP_define}
\Lambda_{\pi}^{\mu}(p',p) &= \gamma^{\mu}F^{(\pi)}_{1Q}(Q^2) + \frac{i\sigma^{\mu\nu}q_{\nu}}{2M}\,F^{(\pi)}_{2Q}(Q^2).
\end{align}
Expressions for these dressed quark form factors are given in App.~\ref{app:pioncloud}. 
The flavor $SU(2)$ piece of the quark-photon vertex (see Eq.~\eqref{eq:barevertex}) including pion loop
effects therefore reads
\begin{align}
\Lambda_{SU(2)}^{\mu,(\pi)}(q) &= Z_{Q} \left(\frac{1}{6}+\frac{\tau_{3}}{2}\right) \gamma^{\mu} \no \\
&\hs{-10mm}
+ \gamma^{\mu} \lf[\frac{1}{2}(1-\tau_3)\, F^{(q)}_{1Q}(Q^2) + \tau_3 \, F^{(\pi)}_{1Q}(Q^2)\right] \no \\
&\hs{-10mm}
+ \frac{i\sigma^{\mu\nu}q_{\nu}}{2M}\lf[\frac{1}{2}(1-\tau_3) \, F^{(q)}_{2Q}(Q^2) + \tau_3 F^{(\pi)}_{2Q}(Q^2)\rg],
\label{M-vertex}
\end{align}
and the corresponding three-flavor vertex is therefore
\begin{align}
\Lambda_{Q}^{\mu,(\pi)}(q) &=  \text{diag}\lf[\Lambda_{SU(2)}^{\mu,(\pi)}(q),~e_s\,\gamma^\mu\rg].
\label{eq:pionvertex}
\end{align}

The $\pi^+$ electromagnetic current, including the effects from the pion cloud, is therefore given by 
Eqs.~\eqref{eq:jpi1}--\eqref{eq:jpi2} with the substitution 
$\Lambda_q^{\mu,\text{(bare)}} \to \Lambda_{Q}^{\mu,(\pi)}(q)$. The $K^+$ 
electromagnetic current, at the same level of sophistication, is obtained via the additional 
substitution $g_\pi\,\tau_{\pm} \to g_K\,\lambda_{\pm}$ in Eqs.~\eqref{eq:jpi1}--\eqref{eq:jpi2}.

\subsection{Pion and Kaon form factors: vector mesons}
\begin{figure}[tbp]
\begin{center}
\includegraphics[width=\columnwidth,clip=true,angle=0]{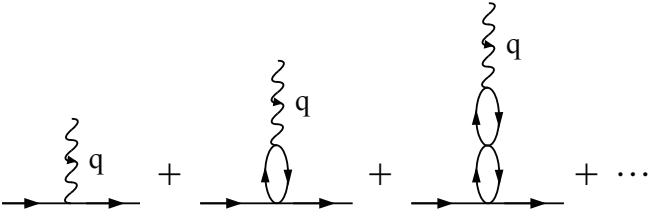}
\caption{Dressing of the quark electromagnetic current from vector mesons.}
\label{VMD}
\end{center}
\end{figure}

The quark-photon vertex receives contributions from the $\bar{q}q$ $T$-matrix in the vector channel, as 
illustrated in Fig.~\ref{VMD}; these contributions are analogous to the familiar vector meson dominance (VMD) 
model~\cite{j1969currents}.
Because of the flavor structure of Eq.~\eqref{NJL lagrangian} the electromagnetic current 
of the light quarks only receives contributions from $\rho^0$ and $\omega$ mesons, while only the $\phi$ meson
couples to the $s$ quark. In this work we will not include the VMD contribution to the 
quark-photon vertex of the $s$ quark because of the larger mass of the $\phi$ meson. 

This is consistent with our earlier approximations of neglecting the contributions of the kaon 
cloud and the mixing between the pseudoscalar ($\pi$) and pseudovector ($a_1$) meson channels.  


Using the transverse Lorentz structure of the bubble diagrams in the vector $\bar{q}q$ channels, the $SU(2)$
piece of the quark-photon vertex becomes
\begin{multline}
\biggl (\frac{1}{6}+\frac{\tau_{3}}{2}\biggr) \gamma^{\mu} \\
\rightarrow \biggl(\frac{1}{6} + \frac{\tau_3}{2} \biggr) \biggl[\gamma^{\mu} - \frac{2\,G_{v}\,\Pi_{v}(q^2)}{1+2\,G_{v}\,\Pi_{v}(q^2)} 
\biggl(\gamma^{\mu}-\frac{\slashed{q}q^{\mu}}{q^2}\biggr)\biggr]\,,
\label{eq:vmd}
\end{multline}
where $\Pi_{v}(q^2)$ is the reduced bubble diagram in the $\rho$ or $\omega$ channel. 
In the on-shell approximation for the external quark momenta the 
$\slashed{q}q^{\mu}$ term in Eq.~\eqref{eq:vmd} does not contribute to the form factors. 
Therefore, the VMD modification of $u$ and $d$ quark-photon vertices is given by
\begin{align}
\gamma^{\mu}\left(\frac{1}{6}+\frac{\tau_{3}}{2}\right) \rightarrow \gamma^{\mu}\left(\frac{1}{6}+\frac{\tau_{3}}{2}\right) 
\frac{1}{1+2G_{v}\Pi_{v}(q^2)}.
\label{vert}
\end{align} 
The quark-photon vertex, including both pion cloud and vector meson effects, is therefore
given by
\begin{align}
\Lambda_{Q}^{\mu}(q) &=  \text{diag}\lf[\Lambda_{SU(2)}^{\mu,(\pi)}(q)\ \frac{1}{1+2G_{v}\Pi_{v}(q^2)},~e_s\,\gamma^\mu\rg].
\label{eq:pionvectorvertex}
\end{align}

VMD effects on the pion form factor can simply be obtained by multiplying the entire form factor
by $\lf[1 + 2\,G_{v}\,\Pi_{v}(q^2)\rg]^{-1}$. For the $K^+$ electromagnetic current 
only to the first term of Eq.~\eqref{eq:barepioncurrent} is multiplied by this factor, because the 
$s$ quark does not couple to the $\omega$ meson.
The form of $\Pi_{v}(q^2)$ is:
\begin{align}
\Pi_{v}(q^2) &= 48i\,q^2\int \frac{d^4 k}{(2\pi)^4} \no \\
&\hs{12mm}
\times\int^{1}_{0} dx\ \frac{x(1-x)}{[k^2-M^2+x(1-x)q^2]^2}.
\end{align}

\section{Results\label{sec:results}}
The NJL model described here depends on two regularization parameters $\Lambda_{UV}$ and 
$\Lambda_{IR}$; the coupling constants $G_\pi$ and $G_v$; and the light ($M$) and strange ($M_s$) dressed quark masses. 

The infrared cut-off simulates one important aspect of confinement and should therefore be similar to $\Lambda_{\text{QCD}}$, 
we choose $\Lambda_{IR} = 0.2\,$GeV. 

The coupling $G_\pi$ and $\Lambda_{UV}$ are fixed by the physical pion 
mass ($m_\pi = 0.140\,$GeV) and pion leptonic decay constant ($f_{\pi}=0.0934\,$GeV); finally $G_v$ and $M_s$ are 
fixed by the physical $\rho$ meson mass ($m_{\omega} \simeq m_{\rho} = 0.776\,$GeV) and physical kaon mass 
($m_K = 0.494\,$GeV). This, therefore leaves one free parameter, the dressed light quark mass $M$, and in this section
we investigate the $M$ dependence of the current quark masses, the kaon decay constant, and the 
pion and kaon form factors.

\begin{table*}[tbp]
\addtolength{\tabcolsep}{4.6pt}
\addtolength{\extrarowheight}{1.2pt}
\begin{tabular}{c|ccccccccccccccc}
\hline\hline
$M$ & $\Lambda_{UV}$ & $G_{\pi}$ & $G_v$ & $M_{s}$ & $m$ & $m_{s}$ & $m_{s}/m$ & $f_K$ & $f_K/f_\pi$ & 
$\left< \bar{\ell} \ell \right>$  & $\left< \bar{s} s \right>$ & $\left< \bar{s} s \right> / \left< \bar{\ell} \ell \right>$ 
\\[0.3ex]
\hline
    0.20 & 1.24 & \hs*{3mm}2.36 & 2.08 & 0.467 & 0.0041 & 0.131 & 31.9 & 0.128 & 1.37 & $-(0.275)^3$ & $-(0.329)^3$  & 1.71 \\
    0.25 & 0.84 & \hs*{3mm}6.12 & 3.06 & 0.502 & 0.0086 & 0.227 & 26.5 & 0.110 & 1.18 & $-(0.214)^3$ & $-(0.224)^3$  & 1.15 \\
    0.30 & 0.71 & 10.6          & 4.52 & 0.540 & 0.0123 & 0.293 & 23.8 & 0.010 & 1.07 & $-(0.190)^3$ & $-(0.180)^3$  & 0.85 \\
    0.35 & 0.66 & 15.0          & 6.64 & 0.573 & 0.0150 & 0.331 & 22.1 & 0.094 & 1.01 & $-(0.177)^3$ & $-(0.159)^3$  & 0.72 \\
    0.40 & 0.64 & 19.3          & 9.60 & 0.609 & 0.0168 & 0.357 & 21.3 & 0.091 & 0.97 & $-(0.170)^3$ & $-(0.148)^3$  & 0.70 \\
\hline\hline
\end{tabular}
\caption{Results for the NJL model parameters: $\Lambda_{UV}$, $G_{\pi}$, $G_v$ and $M_s$; together with
resulting values for the current quark masses, kaon decay constant, and quark condensates, 
all for various values of the dressed light quark mass $M$.
Masses, decay constant and regularization parameters are in units of GeV, the Lagrangian 
couplings, $G_\pi$ and $G_v$, are in units of GeV$^{-2}$, and quark condensates are in units of
GeV$^{3}$.} 
\label{u-mass-P}
\end{table*}

\subsection{Quarks masses and kaon decay constant\label{sec:masses}}
Results for our NJL model parameters; the light ($m$) and strange ($m_s$) current quarks masses;
the kaon decay constant ($f_{K}$); the quark condensates ($\left< \overline{q} q \right>$); 
together with other quantities defined in the text; are 
summarized in Tab.~\ref{u-mass-P}, for values of the dressed light quark mass in the 
range $0.2 \leqslant M  \leqslant 0.4\,$GeV. Empirical analyses of the strange to light current 
quark mass ratio and kaon to pion leptonic decay constant ratio have found 
$m_s/m = 27.5 \pm 1.0$~\cite{Beringer:1900zz} and $f_K/f_{\pi} = 1.197 \pm 0.002 \pm 0.006 \pm 0.001$
~\cite{Beringer:1900zz,Rosner:2013ica}, respectively, 
and a recent QCD analysis \cite{McNeile:2012xh} found  
$\left< \bar{s} s \right> / \left< \bar{\ell} \ell \right> = 1.08 \pm 0.16$ for the
ratio of strange to light ($\ell = u,d$) nonperturbative (physical) quark condensates.

From an inspection of our results presented in Tab.~\ref{u-mass-P} it is clear that good 
agreement with empirical values for the $f_K/f_{\pi}$ and $m_s / m$ ratios, and with the QCD 
analysis for the ratio 
$\left< \bar{s} s \right> / \left< \bar{\ell} \ell \right>$,  is obtained if the
dressed light quark mass has a value near $M \sim 0.25\,$GeV. Therefore, our results favor values 
for $M$ which are considerably lighter than typical values used in effective quark models, like 
the NJL model, where $M \sim 0.4\,$GeV is the norm.

\subsection{Pion and Kaon form factors\label{sec:ffresults}}
\begin{figure}[tbp]
\begin{center}
\includegraphics[width=1.13\columnwidth,clip=true,angle=0]{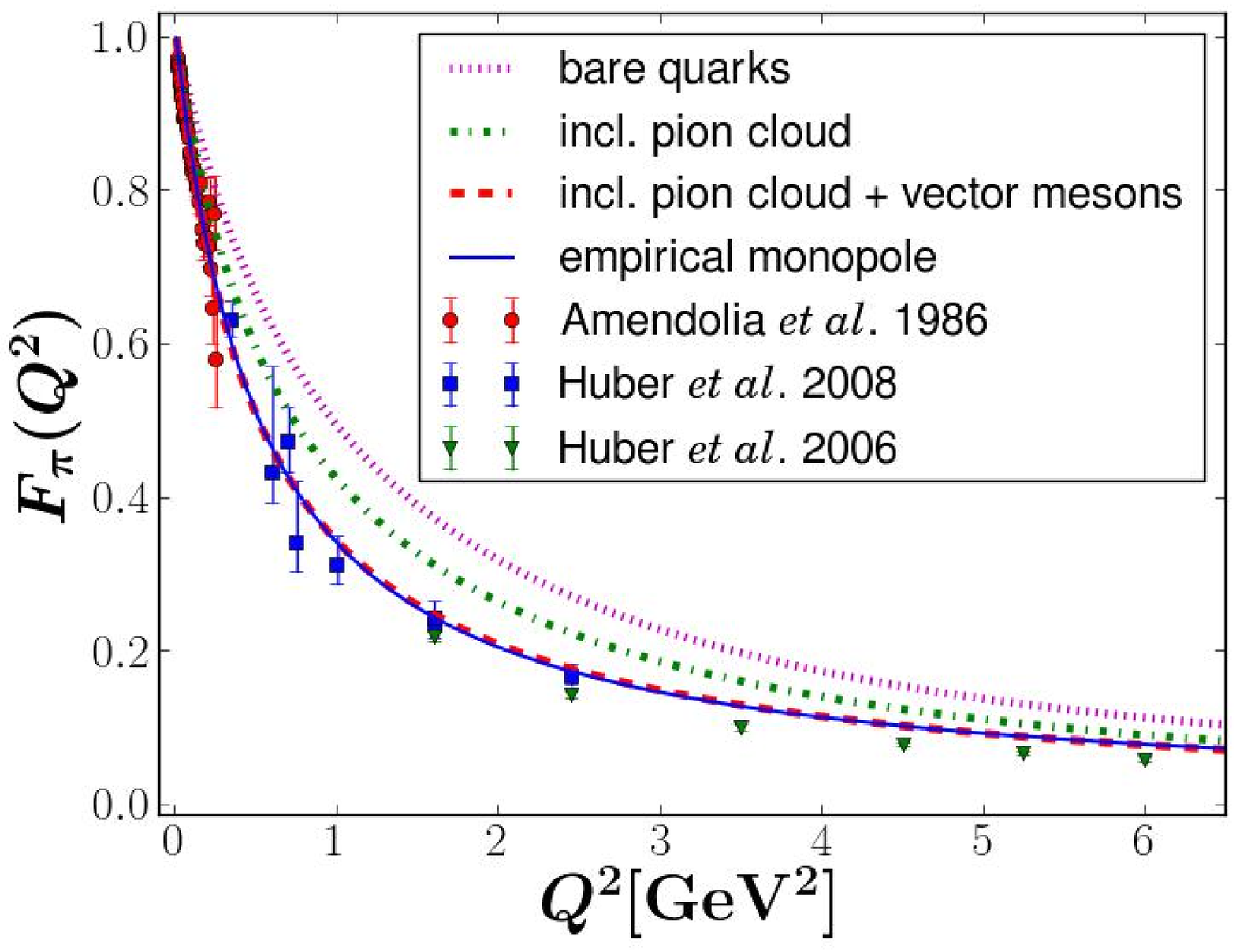}
\caption{(Color online) Pion form factor with $M = 0.25\,$GeV (see Tab.~\ref{u-mass-P}). The data show the experimental values from Amendolia {\it et al.} 1986:~\cite{Amendolia:1986wj}, Huber {\it et al.} 2008:~\cite{Huber:2008id} and projected values from Huber {\it et al.} 2006:~\cite{Huber:2006pac30}.}
\label{Form factor-P2}
\end{center}
\end{figure}

\begin{figure}[tbp]
\begin{center}
\includegraphics[width=1.13\columnwidth,clip=true,angle=0]{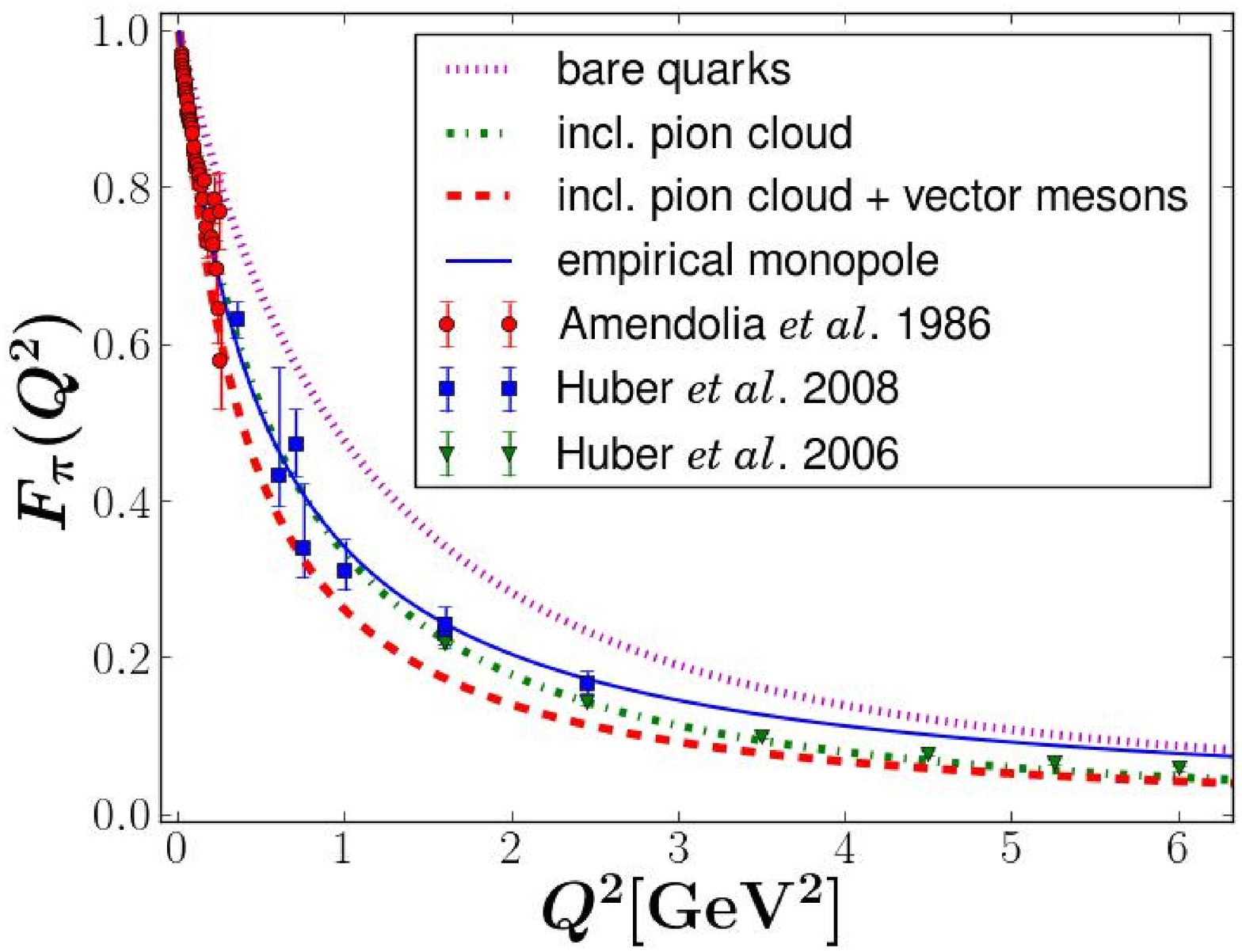}
\caption{(Color online) Pion form factor with $M=0.4\,$GeV (see Tab.~\ref{u-mass-P}). The data show the experimental values from Amendolia {\it et al.} 1986:~\cite{Amendolia:1986wj}, Huber {\it et al.} 2008:~\cite{Huber:2008id} and projected values from Huber {\it et al.} 2006:~\cite{Huber:2006pac30}.}
\label{Form factor-P4}
\end{center}
\end{figure}

Results for the pion form factor are presented in Fig.~\ref{Form factor-P2} for our favored 
value of the dressed light quark mass, namely $M=0.25\,$GeV, and in Fig.~\ref{Form factor-P4}
pion form factor results with $M=0.40\,$GeV are illustrated.
In each figure the dotted line denotes the pion form factor result where the quark-photon 
vertex is treated as point-like (bare); the dash-dotted line includes effects from the pion cloud; 
and the dashed line is the full result which also includes vector mesons in the quark-photon 
vertex. The solid line shows the empirical monopole function 
\begin{align}
F_{\pi}^{(\text{emp})}(Q^2)=\frac{1}{1 + Q^2/0.517\,\text{GeV}^2},
\label{eq:emppionff}
\end{align}
which is constrained to reproduce the central value of the empirical pion charge 
radius $\lf<r_\pi\rg> = 0.672\pm 0.008\,$fm~\cite{Beringer:1900zz}. 
From Fig.~\ref{Form factor-P4} it is clear that the pion form factor with $M=0.4\,$GeV is 
too soft, while the pion form factor result with $M=0.25\,$GeV agrees very well with the empirical result of
Eq.~\eqref{eq:emppionff}.

It is interesting to note that the quark core contributions are rather similar for the $M=0.25\,$GeV and 
$M=0.4\,$GeV cases; and the main difference comes from the pion cloud contributions. This is understood 
by noting that the coupling constant $g_{\pi}$ increases as $M$ becomes larger (see Tab.~\ref{couplings}) -- 
which is consistent 
with the flavor $SU(2)$ quark-level Goldberger-Treiman relation: $M = g_\pi\,f_\pi$~\cite{Klevansky:1992qe} -- 
and therefore leads to larger 
effects from the pion cloud. In addition, as shown in Tab.~\ref{couplings}, the value of $Z_Q$ -- which represents the
probability to find a quark without its pion cloud -- decreases with increasing $M$, leading to
larger pion cloud effects as $M$ increases and to a smaller value of the pion form factor at high $Q^2$, 
because the quark vertex function approaches $Z_Q\,e_q\,\gamma^{\mu}$ as $Q^2 \to \infty$ (see Eq.~\eqref{M-vertex}). 
The end result is that if both the pion cloud and VMD effects are added to the quark core contributions, then the 
data and the empirical monopole function can be reproduced very well for the case $M=0.25\,$GeV, 
while for the case with $M=0.4\,$GeV the calculated form factor is too soft.  

\begin{table}[bp]
\addtolength{\tabcolsep}{17.8pt}
\addtolength{\extrarowheight}{1.2pt}
\begin{tabular}{c|ccc}
\hline\hline
$M~$[GeV] & $g_\pi$  & $g_K$ & $Z_Q$ \\[0.3ex]
\hline
    0.20 & 2.10 & 2.20 & 0.87 \\
    0.25 & 2.62 & 2.79 & 0.85 \\
    0.30 & 3.15 & 3.40 & 0.84 \\
    0.35 & 3.67 & 3.97 & 0.82 \\
    0.40 & 4.20 & 4.55 & 0.80 \\
\hline\hline
\end{tabular}
\caption{Results for the effective quark-meson coupling constants and the quark wave function renormalization, 
for various values of the dressed light quark mass $M$.} 
\label{couplings}
\end{table}

\begin{figure}[tbp]
\begin{center}
\includegraphics[width=1.13\columnwidth,clip=true,angle=0]{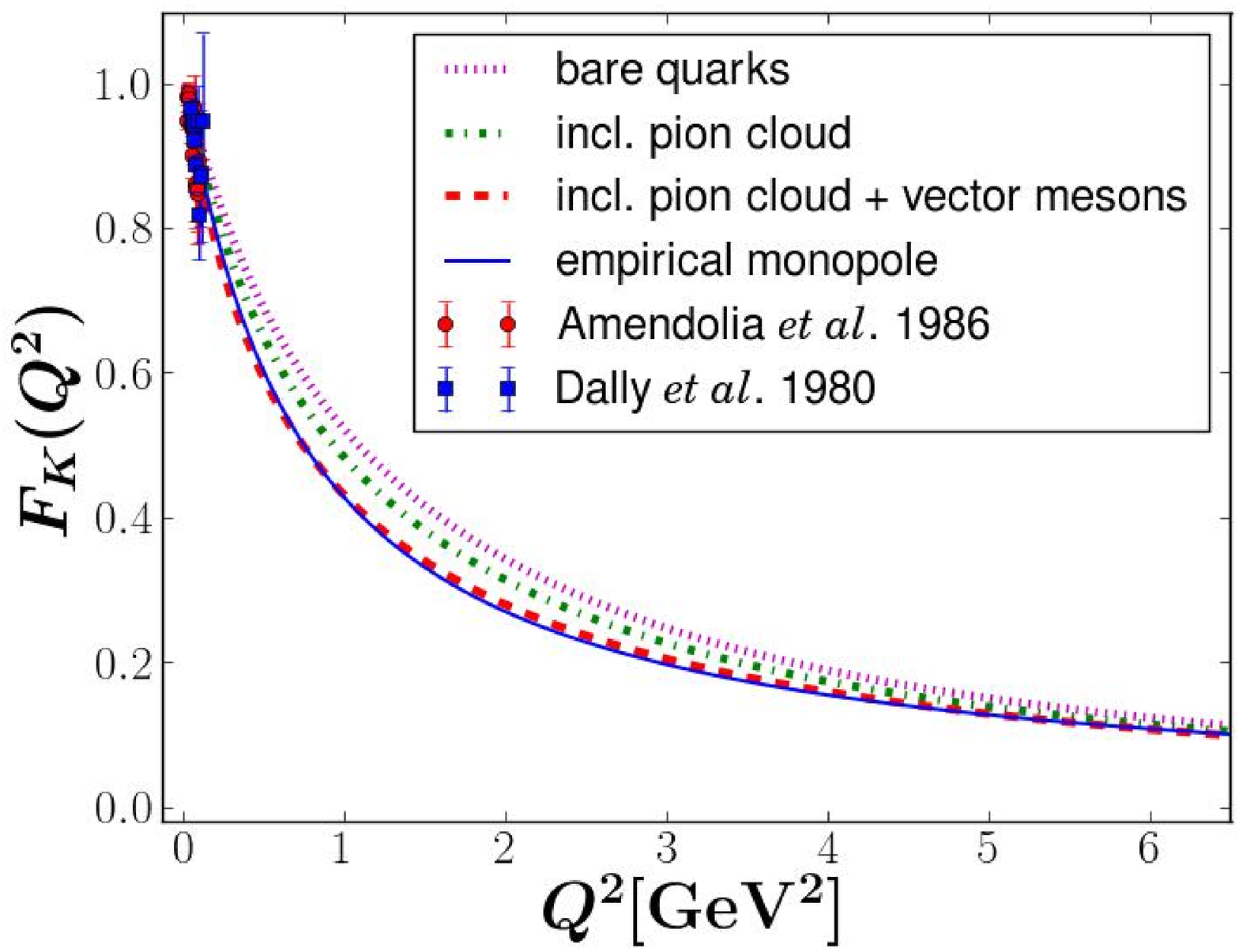}
\caption{(Color online) Kaon form factor with $M=0.25\,$GeV (see Tab.~\ref{u-mass-P}). The experimental values are taken from Amendolia {\it et al.} 1986:~\cite{Amendolia:1986ui} and 
Dally {\it et al.} 1980:~\cite{Dally:1980dj}.}
\label{Form factor-K2}
\end{center}
\end{figure}

\begin{figure}[tbp]                           
\begin{center}                                             
\includegraphics[width=1.13\columnwidth,clip=true,angle=0]{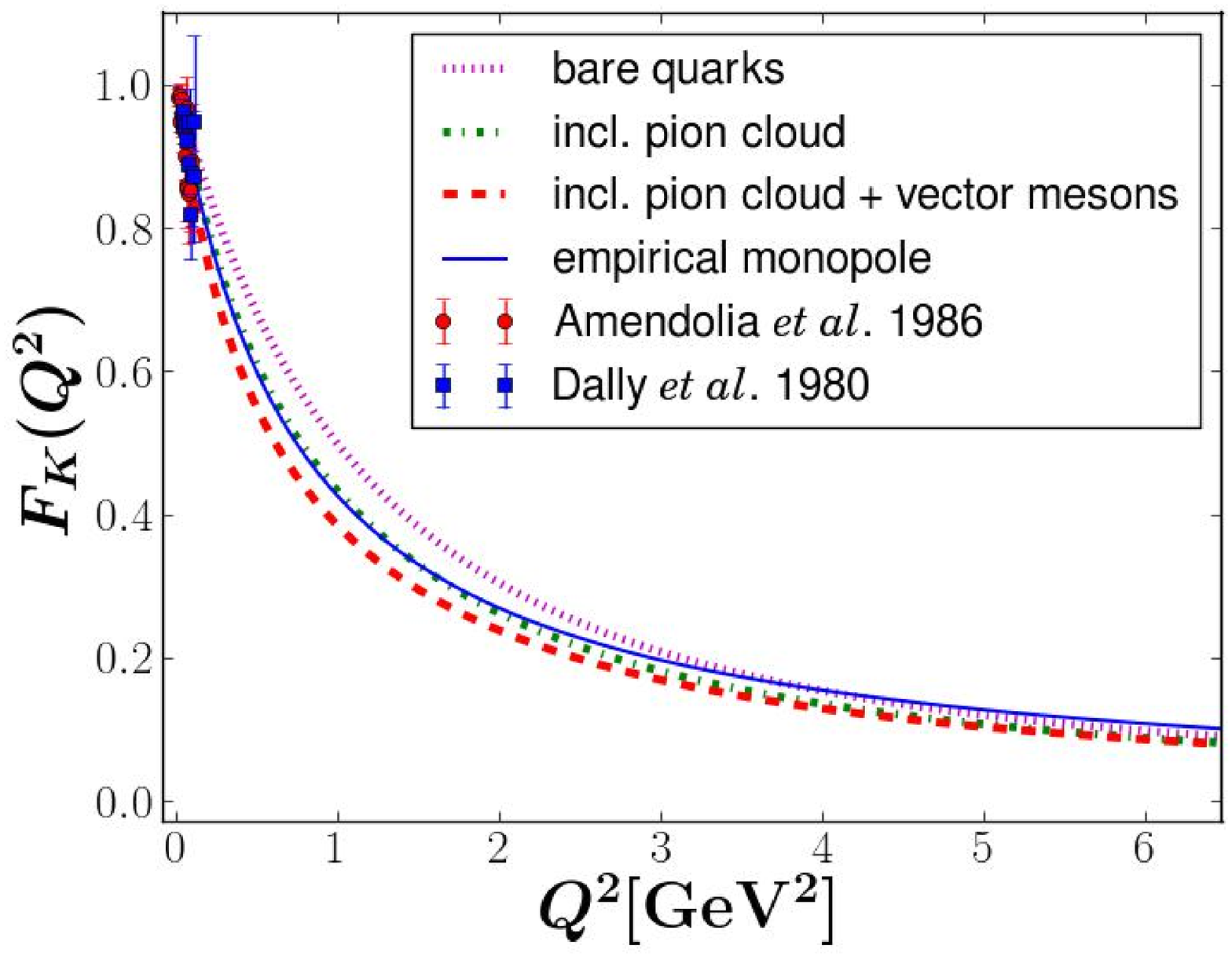}
\caption{(Color online) Kaon form factor with $M=0.4\,$GeV (see Tab.~\ref{u-mass-P}). The experimental values are taken from Amendolia {\it et al.} 1986:~\cite{Amendolia:1986ui} and
Dally {\it et al.} 1980:~\cite{Dally:1980dj}.} 
\label{Form factor-K4}
\end{center}
\end{figure}

Figures~\ref{Form factor-K2} and \ref{Form factor-K4} present kaon form factor results for the 
cases $M=0.25\,$GeV and $M=0.4\,$GeV, respectively. In each figure the dotted line denotes the kaon 
form factor result where the quark-photon vertex is treated as point-like (bare); the dash-dotted 
line includes effects from the pion cloud on the light quark; and the dashed line is the full 
result which also includes vector mesons in the coupling of the photon to the light quark. 
The kaon form factor is poorly known experimentally, however in Figs.~\ref{Form factor-K2} and \ref{Form factor-K4} 
the solid line represents the monopole function:
\begin{align}
F_K^{(\text{emp})}(Q^2)=\frac{1}{1 + Q^2/0.744\,\text{GeV}^2},
\label{eq:empkaonff}
\end{align}
which is constrained to reproduce the central value of the empirical kaon charge radius $\lf<r_K\rg> = 0.560 \pm 0.031\,$fm~\cite{Beringer:1900zz}. The $s$ quark does not couple to the pions or -- under the assumptions used here --
vector mesons, therefore, unlike the pion the kaon form factor is not as sensitive to corrections from the pion cloud and 
vector mesons. However, from Figs.~\ref{Form factor-K2} and \ref{Form factor-K4} it is clear that our kaon form factor
results have better agreement with the empirical result of Eq.~\eqref{eq:empkaonff} when $M=0.25\,$GeV, as opposed to
the case when $M=0.4\,$GeV.

\begin{figure}[tbp]
\begin{center}
\includegraphics[width=1.13\columnwidth,clip]{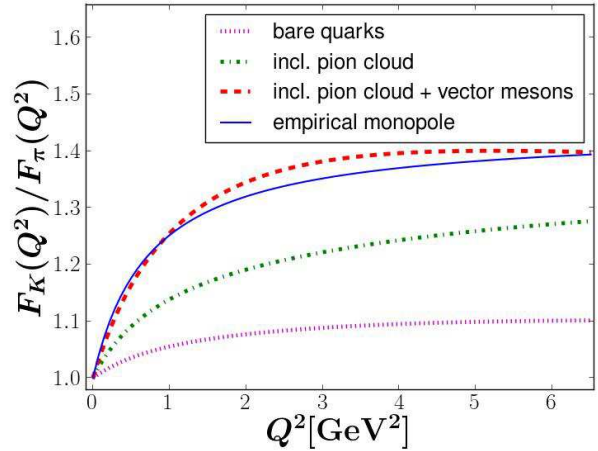}
\caption{(Color online) Results of the ratio of the kaon form factor to the pion form factor $F_{K}(Q^2)/F_{\pi}(Q^2)$ for the case $M=0.25$ GeV.} 
\label{Form factor ratio}
\end{center}
\end{figure}

In Fig.~\ref{Form factor ratio} we present results for the kaon to pion form factor ratio, $F_{K}(Q^2)/F_{\pi}(Q^2)$,
for the case of $M=0.25\,$GeV. We find that this ratio, including effects from the pion cloud and vector mesons, approaches 
$F_{K}/F_{\pi} \sim 1.4$ as $Q^2 \to \infty$. Perturbative QCD predicts that the ratio $F_{K}/F_{\pi}$ should approach 
$f_{K}^2/f_{\pi}^2$ as $Q^2 \rightarrow \infty$~\cite{Farrar:1979aw,Lepage:1980fj}. 
Since our calculation for $M=0.25$ GeV reproduces the experimental values
for both decay constants with the squared ratio $f_K^2/f_{\pi}^2 = 1.4$ (see Tab.~\ref{u-mass-P}), 
we can say that our NJL model result for $M=0.25\,$GeV is consistent with the prediction based on perturbative QCD.
This agreement cannot be attained for the case
of $M=0.4\,$GeV, where our calculated ratio of form factors becomes larger than the calculated ratio of decay constants.  

However, before drawing firm conclusions about the behavior of the form factors for large values of
$Q^2$, one should take into account the contributions of the mixing between the 
pseudoscalar ($\pi$) and pseudovector ($a_1$) meson channels, as mentioned at the end of Sect.\,\ref{mesons couplings}. 



\begin{table*}[tbp]
\addtolength{\tabcolsep}{4.3pt}
\addtolength{\extrarowheight}{1.9pt}
\begin{tabular}{c|ccc|ccc|cccccc}
\hline\hline
$M$ & $\lf< r_\pi \rg>^{\text{(bare)}}$ & $\lf< r_K \rg>^{\text{(bare)}}$ & $\lf< r_\pi \rg>^{\text{(bare)}}/\lf< r_K \rg>^{\text{(bare)}}$ 
    & $\lf< r_\pi \rg>^{(\pi)}$       & $\lf< r_K \rg>^{(\pi)}$       & $\lf< r_\pi \rg>^{(\pi)}/\lf< r_K \rg>^{(\pi)}$ 
    & $\lf< r_\pi \rg>$             & $\lf< r_K \rg>$              & $\lf< r_\pi \rg>/\lf< r_K \rg>$ \\[0.3ex]
\hline
    0.20 & 0.455 & 0.430 & 1.06  &  0.527 & 0.481 & 1.09  &  0.645 & 0.571 & 1.13\\
    0.25 & 0.489 & 0.465 & 1.05  &  0.589 & 0.530 & 1.11  &  0.690 & 0.608 & 1.14\\
    0.30 & 0.497 & 0.474 & 1.05  &  0.627 & 0.553 & 1.13  &  0.724 & 0.630 & 1.15\\
    0.35 & 0.488 & 0.468 & 1.04  &  0.649 & 0.562 & 1.15  &  0.750 & 0.643 & 1.17\\
    0.40 & 0.472 & 0.453 & 1.04  &  0.663 & 0.563 & 1.18  &  0.773 & 0.653 & 1.18\\
\hline\hline
\end{tabular}
\caption{Pion and kaon charge radii (in units of fm) and their ratios for various 
choices of the dressed light quark mass (in GeV). The case labeled by (bare) 
corresponds to dressed quarks treated as point-like (see Eq.~\eqref{eq:barevertex}); 
the case labeled by ($\pi$) corresponds to including pion cloud effects (see Eq.~\eqref{eq:pionvertex});
and the case with no superscript corresponds to including effects from both the pion cloud and vector mesons (see Eq.~\eqref{eq:pionvectorvertex}).
The values for the NJL model parameters needed to obtain these results are given in Tab.~\ref{u-mass-P}.}
\label{tab:radii}
\end{table*}

\subsection{Pion and kaon charge radii}
The charge radius, $\lf<r_k\rg>$, of the pion and kaon is obtained from the corresponding form factor 
via the relation:
\begin{align}
\lf<r_k\rg> = \sqrt{\lf.-6\frac{\partial F_k(Q^2)}{\partial Q^2}\rg|_{Q^2=0}}.
\end{align}
Our results are given in Tab.~\ref{tab:radii} for the three variations of the photon coupling to
the dressed quarks. For the case where the quark-photon vertex is treated as point-like (bare), 
the charge radii of the pion and kaon (including their ratios) do not depend strongly on the 
dressed $u$ and $d$ quark mass. Ref.~\cite{Beringer:1900zz} gives empirical values for the 
pion and kaon charge radii of: $\lf<r_\pi\rg> = 0.672\pm 0.008\,$fm and $\lf<r_K\rg> = 0.560 \pm 0.031\,$fm,
with the ratio therefore equal to $\lf<r_\pi\rg>/\lf<r_K\rg> = 1.20 \pm 0.08\,$. For
the bare quark-photon coupling we therefore find that the pion and kaon charge radii, 
together with their ratio, are too small. 

Results for the pion and kaon charge radii, including effects form the pion cloud around the
dressed quarks, are presented in the third sector of Tab.~\ref{tab:radii}. The pion cloud leads 
to a considerable enhancement of the pion charge radius, and a less pronounced enhancement of the 
kaon charge radius, bringing all results into better agreement with the empirical values. 
As discussed earlier, increasing the dressed quark mass $M$, results in larger values for
$g_\pi$ and smaller values for $Z_{Q}$ (see Tab.~\ref{tab:radii}), and both of these effects
increase the pion cloud effects for the charge radii and form factors.

Results for pion and kaon charge radii which include effects from vector mesons and the pion cloud are presented
in the final sector of Tab.~\ref{tab:radii}. Good agreement with the empirical results is obtained 
when the dressed light quark mass is in the range $0.2 \leqslant M \leqslant 0.25\,$GeV; while
for large dressed $u$ and $d$ quark masses the charge radii are too large.


Within our present model description, 
we have therefore found that the electromagnetic properties of the pion and kaon, that is, their
charge radii and the $Q^2$ dependence of their form factors (see Sect.~\ref{sec:ffresults}), are 
described very well if $M = 0.25\,$GeV. 

This is consistent with the observations discussed in
Sect.~\ref{sec:masses} where results in good agreement with experiment and QCD based analyses are 
obtained for the kaon decay constant, current quark masses and quark condensates (see Tab.~\ref{u-mass-P}) 
if the dressed light quark mass is approximately $M \simeq 0.25\,$GeV.

\section{Summary\label{sec:summary}}
The NJL model, including effects of the pion cloud and vector mesons at the quark level, has been 
used to study the pion and kaon electromagnetic form factors. An important motivation for this study
was to investigate the dressed light quark ($u$ and $d$) mass dependence of pion and kaon
observables. 

We began with results for the $u,\,d,\,s$ current quark masses, quark condensates, and the kaon decay constant. 
Within the limits of our approximation scheme, we found that the results for the current quark mass ratio 
$m_s/m$, the ratio of condensates $\left< \bar{s}s \right> / \left< \bar{\ell} \ell \right>$, and the kaon decay 
constant are in good agreement with empirical and QCD based results if our dressed
light quark mass is approximately $M \sim 0.25\,$GeV. 

We next studied the dressed $u$ and $d$ quark mass 
dependence of the pion and kaon form factors.  We found that the pion cloud and vector mesons 
have a substantial effect on these form factors, and that pion cloud effects
increase as the dressed light quark mass becomes larger (with fixed pion mass), as
a consequence of the increased pion--quark-quark coupling. 
One important effect of the pion cloud is to
enhance the pion charge radius more than the kaon charge radius, bringing the charge radii as well as their
ratio in better agreement with empirical results. 

We found that, within the limits of our approximation scheme, the available data on the form factors
and charge radii are well described with relatively small values for the dressed $u$ and $d$ quark mass
of approximately $M \sim 0.25\,$GeV. 

For this case, we also found that the ratio of the kaon to pion form factor 
for large values of $Q^2$ agrees very well with the perturbative QCD prediction.  

Our finding that a dressed $u$ and $d$ quark mass of $M \sim 0.25\,$GeV leads to a good description of
the pion and kaon electromagnetic properties, the kaon decay constant, and to reasonable values for
the current quark masses, quark condensates and their ratios, is interesting, because so far calculations in 
constituent-like quark models, 
e.g., the NJL model or chiral soliton models~\cite{Kubota:1999hx}, mostly use $u$ and $d$ quark masses in the range 
$0.3 \lesssim M \lesssim 0.4\,$GeV. 


We emphasize that, because the infrared cut-off in our calculation eliminates unphysical thresholds 
for the decay of hadrons into quarks, there is no inherent problem with describing
the heavier hadrons by using smaller dressed light quark masses. 

For example, in our calculation
a vector meson mass of $0.776\,$GeV is easily obtained, we also confirmed that a
nucleon mass of $0.94\,$GeV can be reproduced with reasonable parameters.\footnote{In the simplest quark - scalar diquark model for the nucleon, used for example in 
Ref.~\cite{Bentz:2001vc}, one finds that for $M=0.25\,$GeV the experimental nucleon mass can be 
reproduced by using $G_s/G_{\pi}=0.56$, 
where $G_s$ is the 4-Fermi coupling constant in the scalar quark-quark channel.} 
It would be interesting to explore other
hadronic properties, e.g., the nucleon electromagnetic form factors, in the domain of smaller dressed
quark masses in this model description. 

\section*{Acknowledgements}
Y.N. wishes to thank E.V. Morooka and S. Hakamada for their careful reading of the manuscript and for their
help in the preparation of some figures of the paper. W.B. acknowledges support by the Grant in Aid for 
Scientific Research (Kakenhi) of the Japanese Ministry of Education, Culture, Sports, Science and Technology, 
project no. 20168769.
I.C. is supported by the Department of Energy, Office of Nuclear Physics, contract no. DE-AC02-06CH11357.

\appendix
\section{Regularization method \label{app:pt}}
To evaluate 4-dimensional integrals, we first introduce Feynman parametrization and perform shifts of the
loop momentum so that the integrand depends only on $k^2$, where $k$ is the loop momentum (plus
other fixed parameters). We then perform a Wick rotation and use 4-dimensional spherical coordinates to obtain
\begin{align}
\int d^4 k\ f(k^2) = 2\pi^{2}i \int^{\infty}_{0} dk_{E}\, k_{E}^3\ f(-k_{E}^2),
\end{align}
where $k_{E} = \sqrt{k_{0}^2 + \vec{k}^2}$ is the Euclidean length. Next, we consider the following identity:
\begin{align}
\frac{1}{D^n} = \frac{1}{(n-1)!}\int^{\infty}_{0}d\tau\,\tau^{n-1}\,e^{-\tau D},
\end{align}
where $D$ is the denominator of the integral. Here the cut-off parameters $\Lambda_{UV}$ and $\Lambda_{IR}$ are
introduced as follows:
\begin{align}
&\frac{1}{(n-1)!}\int^{\infty}_{0}d\tau\ \tau^{n-1}\,e^{-\tau D} \no \\ 
&\hs{20mm}
\longrightarrow
\frac{1}{(n-1)!}\int^{1/\Lambda_{IR}^2}_{1/\Lambda_{UV}^2} d\tau\ \tau^{n-1}\,e^{-\tau D}.
\end{align}
Only the ultraviolet cut-off parameter, $\Lambda_{UV}$, is needed to make the integrals finite, however
including an infrared cut-off, $\Lambda_{IR}$, eliminates unphysical thresholds for the decay of hadrons
into quarks and plays the role of simulating the confinement in the NJL model. Therefore, in the case 
of the loop integrals for quarks the infrared cut-off should satisfy $\Lambda_{IR} \sim \Lambda_{\text{QCD}}$,
however for loop integrals involving virtual pions, where these pions should not be confined, we set $\Lambda_{IR}=0$.

\section{Formulae for the bubble diagrams \label{app:bubble}}
In this Appendix we give formulae for the regularized bubble diagrams for the pion and kaon, which 
enter the pole condition equations, and the coupling constants of pion and kaon to quarks. These 
bubble diagrams take the form
\begin{align}
\Pi_{\pi}(p^2) &= 12i \int^{1}_{0} dx \int \frac{d^4 k}{(2\pi)^4}\no \\
 \times &\left[\frac{p^2}{[k^2+p^2x(1-x)-M^2]}-\frac{2}{k^2-M^2} \right],
\label{pipion}
\end{align}
\begin{align}
\Pi_K(p^2) &= 12i \int^{1}_{0} dx \int \frac{d^4 k}{(2\pi)^4} \no \\
 &\times \left[\frac{p^2-(M_s-M)^2}{[k^2+p^2x(1-x)-x(M^2-M_s^2)-M_s^2]} \no \right. \no \\
 &\left.-\frac{1}{k^2-M_s^2}-\frac{1}{k^2-M^2} \right].
\label{pikaon}
\end{align}
Introducing the cut-off parameters as explained in App.\,\ref{app:pt}, the regularized bubble diagrams are given by
\begin{align}
\Pi_{\pi}(p^2) &= -\frac{3}{4\pi^2}\int^{1}_{0}dx \int^{1/\Lambda_{IR}^2}_{1/\Lambda_{UV}^2} d\tau\ \frac{1}{\tau} \no \\
&\hs{7mm}
\times \left[\frac{2}{\tau}\, e^{-\tau M^2} + p^2\,e^{-\tau [M^2 - x(1-x)\,p^2]}\right], \\[0.3ex]
\Pi_{K}(p^2) &= -\frac{3}{4\pi^2}\int^{1}_{0}dx \int^{1/\Lambda_{IR}^2}_{1/\Lambda_{UV}^2} d\tau\ \frac{1}{\tau} \no \\
&\hs{-12mm}
\times \left[\lf[p^2-(M_{s}-M)^2\rg] e^{-\tau [M_{s}^2+x(M^2-M_{s}^2) - x(1-x)\,p^2]}\right. \no \\
&\hs{26mm}
\left. + \frac{1}{\tau} \lf[e^{-\tau M^2}+e^{-\tau M_{s}^2}\rg]\right].
\end{align}



\section{Renormalization of pion cloud effects~\label{app:renormalize}}
\begin{figure}[tbp]
\begin{center}
\includegraphics[scale=0.9]{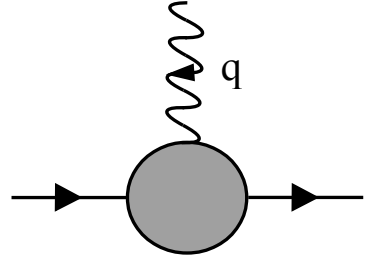}
\caption{Graphical representation of the general quark electromagnetic vertex $\Gamma^{\mu}$.}
\label{general vertex}
\end{center}
\end{figure}

In this Appendix the standard techniques of perturbative renormalization are briefly reviewed. These
techniques are applied to the renormalization of the mass, wave function normalization and charge
of a dressed quark from a pion cloud (see Figs.~\ref{PC-figure} and \ref{SelfEnergy}), thereby giving the renormalized (``physical'') values.
We will restrict the discussion in this Appendix to the flavor $SU(2)$ case, 
because the strange quark cannot couple to the pion due to isospin conservation.
Further, we will assume isospin symmetry ($m_u = m_d = m$) and refer only to 
the scalar and pseudoscalar interaction terms of the Lagrangian given in Eq.~\eqref{NJL lagrangian}.
Labeling the unrenormalized quantities with a subscript $0$, and
including explicitly the coupling to an external vector field $V^{\mu}$, we have:
\begin{align}
{\cal L} &= \overline{\psi}_0 \left( i \slashed{\partial} - m_{ 0} \right) \psi_0 
- \left( \overline{\psi}_0 \gamma^{\mu} {e}_{ 0} \, \psi_0 \right) V_{\mu} \no \\
&\hs{20mm}+ G_{\pi 0} \left[ \left(\overline{\psi}_0 \psi_0 \right)^2 - 
\left(\overline{\psi}_0 \gamma_5\tau_i  \psi_0\right)^2 \right],   
\label{one}
\end{align}
where $\psi = (u,\,d)$ and ${e}_{0}$ is the unrenormalized flavor $SU(2)$ quark charge in units of the 
elementary charge.  
The renormalized quantities -- which are the same as in the main
text -- are introduced by the scale transformations
\begin{align}
\psi_0 &= \sqrt{Z_Q}\, \psi, &
{m}_0 &= \frac{{m}}{Z_Q}, &
G_{\pi 0} &= \frac{G_{\pi}}{Z_Q^2}, &
{e}_{0} &= \frac{Z_{V}}{Z_Q} {e},
\label{two}
\end{align}
where $Z_{V}$ is the quark vertex renormalization for an external vector field,
defined at zero momentum transfer. As usual, gauge invariance leads to the
Ward identity result $Z_{V} = Z_Q$, so that the electric charge is not
renormalized and given by $\left(\frac{1}{6} + \frac{\tau_3}{2} \right)$ as in
Eq.~\eqref{eq:barevertex}. (Here we do not consider the renormalization of the external
vector field.) After the scale transformation the Lagrangian of Eq.~\eqref{one} becomes   
\begin{align}
{\cal L} &= \overline{\psi} \left( Z_Q \, i \slashed{\partial} - {m} \right) \psi 
- Z_{V} \left( \overline{\psi} \gamma^{\mu} {e} \, \psi \right) V_{\mu}  \no \\
&\hs{29mm}+ G_{\pi} \left[ \left(\overline{\psi} \psi \right)^2 - 
\left(\overline{\psi} \tau_i \gamma_5 \psi\right)^2 \right].   
\label{three}
\end{align}
The mass renormalization is performed in the usual manner, that is, by adding and subtracting the term
$-\overline{\psi} \left(M - m \right) \psi$, where the subtracted term is treated as a counter term:
\begin{align}
{\cal L} &= \overline{\psi} \left(Z_Q \, i \slashed{\partial} - {M} \right) \psi 
- Z_{V} \left( \overline{\psi} \gamma^{\mu} {e} \, \psi \right) V_{\mu} \no \\ 
&\hs{4mm}+ G_{\pi} \left[ \left(\overline{\psi} \psi \right)^2 - 
\left(\overline{\psi} \tau_i \gamma_5 \psi\right)^2 \right]  
+ \overline{\psi} \left(M - m \right) \psi.
\label{four}
\end{align}
Following the standard procedure, we split $\overline{\psi} \psi$ in the second line of Eq.~\eqref{four}
into an expectation value in the constituent quark vacuum, 
and a normal ordered product, which by
definition has no vacuum expectation value. Inserting $\overline{\psi} \psi = \langle \overline{\psi} \psi \rangle
\,\,+ :\!\!\overline{\psi} \psi\!\!:$ into the second line of Eq.~\eqref{four}, and requiring that the result becomes a
``true'' residual interaction without terms linear in $:\!\!\overline{\psi} \psi\!\!:$, we obtain
the familiar gap equation
\begin{align}
M &= m - 2\, G_{\pi}\, \langle \overline{\psi} \psi \rangle \no \\
&= m + 48 i\, M\, G_{\pi} \int \frac{d^4 k}{\left(2 \pi \right)^4}
\ \frac{1}{k^2 - M^2 + i \epsilon}.   
\label{five}
\end{align}
For the isospin symmetric flavor $SU(2)$ case this is the same as Eq.~\eqref{constituent quark mass}. 
The gap equation can therefore be viewed as a definition of normal ordering and the constituent quark vacuum.
Any contribution to the mass shift, for example, from the virtual pion cloud around the 
dressed quark (see Fig.~\ref{SelfEnergy}),
must also be included in the counter term proportional to
$(M-m)$ in Eq.~\eqref{four}, which just leads to a redefinition of normal ordering and the dressed
quark vacuum \cite{Nambu:1961tp,Nambu:1961fr}. 
The Lagrangian therefore becomes 
\begin{align}
{\cal L} &= \overline{\psi} \left(Z_Q \, i \slashed{\partial} - {M} \right) \psi 
- Z_{V} \left( \overline{\psi} \gamma^{\mu} {e} \, \psi \right) V_{\mu} \nonumber \\
&\hs{22mm}
+ G_{\pi} \left[ \left(:\!\overline{\psi} \psi\!: \right)^2 - 
\left(:\!\overline{\psi}\gamma_5 \tau_i  \psi\!:\right)^2 \right],  
\label{five1}
\end{align}
where an irrelevant constant ($c$-number) term has been dropped. 
The quark wave function renormalization factor $Z_Q$ is determined perturbatively
from the requirement that the dressed quark propagator, including the self energy term illustrated in
Fig.~\ref{SelfEnergy}, becomes $S(p) = 1/\left(\slashed{p} - M + i \epsilon \right)$ as $\slashed{p} \rightarrow M$
[see Eq.~\eqref{propagator}]. This gives 
\begin{align}
Z_Q = 1 + \lf.\frac{\partial \Sigma(p)}{\partial \slashed{p}}\rg|_{\slashed{p} = M} \,,
\end{align}
which is just Eq.~\eqref{renormalization constant}. Therefore, as long as pion cloud effects are
only included on the level of the mass and wave function renormalization of the dressed quark, there is no 
change in the standard NJL model description. 
To demonstrate this in more detail we verify various low-energy theorems which are
important herein:
\begin{itemize}
\item {\bf Goldstone theorem}: By using Eq.~\eqref{five} and 
the form of the bubble graph given by
Eq.~\eqref{pipion}, it is easy to verify the identity 
\begin{align}
\langle \overline{\psi} \psi \rangle = M \, \Pi_{\pi}(0),
\label{six}
\end{align}
which relates the quark condensate and the bubble graph $\Pi_{\pi}(p^2)$ at $p^2=0$. It then follows from the
gap equation [Eq.~\eqref{five}] and the pion pole condition of Eq.~\eqref{pipole} that $m_{\pi}^2 = 0$ if $m=0$.
\item {\bf Goldberger-Treiman (GT) relation (at the quark level)}:
Let us write the expression for the pion decay constant, which is obtained from Eqs.~\eqref{eq:kaon decay} and \eqref{eq:kaon decay 1}, by the substitutions $M_s \rightarrow M$, $g_K \rightarrow g_{\pi}$ and $m_K \rightarrow m_{\pi}$, as
\begin{align}
f_{\pi} = g_{\pi}\, M\, I(p^2= m_{\pi}^2), 
\label{seven}
\end{align}
where the function $I(p^2)$ is defined by
\begin{align}
\hs{6mm}I(p^2) = -12 i\int \frac{d^4k}{(2 \pi)^4}
\frac{1}{\left[(p+k)^2 - M^2 \right] \left[k^2 - M^2\right]}.
\label{eight}
\end{align} 
This function is related to the bubble graph $\Pi_{\pi}(p^2)$ as follows [see Eq.~\eqref{pipion}]:
\begin{align}
\Pi_{\pi}(p^2) - \Pi_{\pi}(0) = - p^2 I(p^2).
\label{nine}
\end{align}
Using the derivative of this relation w.r.t. $p^2$ and also Eq.~\eqref{kaon quark constant}, 
it follows that Eq.~\eqref{seven} can be written as
\begin{align}
M = g_{\pi} f_{\pi} \left(1 + C \right).
\label{gt}
\end{align}
Here $C$ is defined as
\begin{align}
C = m_{\pi}^2 \, \frac{I'(m_{\pi}^2)}{I(m_{\pi}^2)},
\end{align}
where the prime denotes differentiation w.r.t. $p^2$.
Eq.~\eqref{gt} is the GT relation at the quark level in the present context, 
where $g_{\pi}$ and $f_{\pi}$ are defined at
the pion pole. We also note that in Eq.~\eqref{seven} we assumed that the axial coupling constant
of the dressed quark is given by its bare value, equal to unity. 
If we would use instead a model value for $g_A$, which may be calculated for example from the pion cloud 
similarly to Fig. \ref{PC-figure},    
in the pion decay diagram of Fig.\,\ref{decay constant}, then Eq.~\eqref{seven}
gets a factor $g_A$ on the r.h.s., and the GT relation [Eq.~\eqref{gt}] takes the familiar
form $M\,g_A = g_{\pi}\,f_{\pi}$ in the chiral limit ($m_{\pi}^2 \rightarrow 0$).
\item
{\bf Gell-Mann--Oakes--Renner (GOR) relation}: 
Note that the above relations allow us to express the gap equation and the
pion pole condition in terms of the bubble graph as follows:
\begin{align}
\hs*{11mm}1 + 2 G_{\pi} \Pi_{\pi}(0) &= \frac{m}{M},  \\
\Pi_{\pi}(m_{\pi}^2) - \Pi_{\pi}(0) &= - m_{\pi}^2 I(m_{\pi}^2) = 
- \frac{m}{2 G_{\pi} M}. 
\end{align}
The GOR relation is then obtained as follows:
\begin{align}
- m\,\langle \overline{\psi} \psi \rangle &= \frac{M\,m}{2\,G_{\pi}} \left( 1 - \frac{m}{M} \right) \no \\
&=  M^2\, m_{\pi}^2\, I(m_{\pi}^2)\left( 1 - \frac{m}{M} \right) \no\\
&=  m_{\pi}^2\,  f_{\pi}^2 \left( 1 + C \right) \left(1 - \frac{m}{M} \right),
\label{gor}
\end{align}
where we have used Eqs.~\eqref{seven} and \eqref{gt} to obtain the last line. In the chiral limit Eq.~\eqref{gor}
becomes the familiar GOR relation.
\end{itemize}
This concludes the verification of the low energy theorems in our present context.
Finally we return to the Lagrangian of Eq.~\eqref{five1} and discuss the treatment
of the quark electromagnetic vertex $\Gamma^{\mu}$, which is represented generally by Fig.~\ref{general vertex}.
The ``bare'' vertex is given by $Z_{V}\, {e}\, \gamma^{\mu} = Z_{Q}\, {e}\, \gamma^{\mu}  $ and 
renormalization in a ``global'' sense would simply mean charge renormalization, that is,
according to the definition of $Z_{V}$, the replacement
$\gamma^{\mu} \rightarrow \frac{1}{Z_{V}}\, \gamma^{\mu}$. This would give the
renormalized quark vertex as $\Gamma^{\mu} = {e} \gamma^{\mu}$, which is
correct in the limit $q \rightarrow 0$.
One of the main interests of our present work, however,
is to resolve this electromagnetic vertex on the level of the virtual pion cloud. 
For this purpose, the bare vertex 
$Z_{Q}\, {e}\, \gamma^{\mu}$, which includes the counter term from wave function renormalization,
is supplemented by the
corrections due to the virtual pion cloud, as shown in Fig.~\ref{PC-figure}.
In the pion loop diagrams (second and third diagrams of Fig.~\ref{PC-figure}), we
do not attempt to further resolve the pion cloud around the dressed quark.
Therefore, by using $\Gamma^{\mu} = {e} \gamma^{\mu}$ at the quark-photon vertex in those diagrams, 
we obtain the expressions
given in Eqs.~\eqref{Q-formfactor-temporary} and \eqref{Pion-photon-C}. 
Further, inclusion of the VMD contributions (see Fig.~\ref{VMD}) leads to the correction factor
given in Eq.~\eqref{vert}. 


\section{Formula for pion cloud effects~\label{app:pioncloud}}
To calculate the quark wave function renormalization constant $Z_Q$ arising from the pion cloud, 
we need the derivative of dressed $u$ and $d$ quark self-energy with respect to $\slashed{p}$,
that is 
\begin{align}	
&\lf.\frac{\partial \Sigma (p)}{\partial \slashed{p}}\rg|_{\slashed{p}=M}
= \frac{3\,g_{\pi}^2}{8\pi^2}\int^{1}_{0}dx \int^{\infty}_{1/\Lambda_{UV}^{2}} d \tau \no \\
&\hs{8mm}
\times\left[x(1-x)^2M^2 - \frac{x}{2\tau} \right]e^{-\tau[(1-x)^2M^2 + x\,m_{\pi}^2]}.
\end{align}

In the following we give the formulae for the functions related to the quark electromagnetic 
vertex corrections arising from the pion cloud: 
\begin{align}
F_{1Q}^{(q)}(Q^2) &= \frac{g_{\pi}^2}{32\pi^2}
\biggl\{\int^{1}_{0}dx \int^{\infty}_{1/\Lambda_{UV}^{2}}\!\! d\tau\, \frac{2}{\tau}\, e^{-\tau[x(1-x)Q^2+M^2]} \no \\
&\hs{-13mm}
+ \int^{1}_{0}dx \int^{x}_{-x}dy \int^{\infty}_{1/\Lambda_{UV}^{2}}\!\! d\tau \left[2x^2M^2 - m_{\pi}^2 -\frac{1}{\tau}\right]
e^{-\tau\,A} \biggr\}, \\
F_{2Q}^{(q)}(Q^2) &= -\frac{g_{\pi}^2\,M^2}{16\pi^2}
\int^{1}_{0}dx \int^{x}_{-x}dy \int^{\infty}_{1/\Lambda_{UV}^{2}}\!\! d\tau\ x^2\, e^{-\tau\,A}, \\
F_{1Q}^{(\pi)}(Q^2) &= F_{\pi}^{(\text{bare})}(Q^2)\,\frac{g_{\pi}^2}{16\pi^2} \no \\
&\hs{-14mm}
\times \int^{1}_{0}dx \int^{x}_{-x}dy \int^{\infty}_{1/\Lambda_{UV}^{2}} d\tau \left[\frac{1}{\tau} - 2\,(1-x)^2\,M^2\right]e^{-\tau\,B}, \\
F_{2Q}^{(\pi)}(Q^2) &= F_{\pi}^{(\text{bare})}(Q^2)\,\frac{g_{\pi}^2\,M^2}{8\pi^2}  \no \\
&\hs{-2mm}
\times \int^{1}_{0}dx \int^{x}_{-x}dy \int^{\infty}_{1/\Lambda_{UV}^{2}} d\tau\ (1-x)^2\,e^{-\tau\,B}.
\end{align}
where $A = (1-x)m_{\pi}^2 + x^2 M^2 + \frac{1}{4}\,(x^2-y^2)\,Q^2$ and $B = x\,m_{\pi}^2 + (1-x)^2 M^2 + \frac{1}{4}\,(x^2-y^2)\,Q^2$.
The above expressions are used in Eq.~\eqref{M-vertex} when including pion cloud contributions to the pion and kaon form factors.

The contribution of the loop calculation for the term proportional to $\gamma^{\mu}$, 
in Eq.~\eqref{M-vertex}, to the pion form factor is simply proportional to
the pion form factor with bare quark-photon coupling (see Eq.~\eqref{P-form factor}). 
Similarly, the contribution to the kaon form factor is proportional to the sum of 
the first term and third term of Eqs.~\eqref{K-form factor}.

When using the quark-photon vertex of Eq.~\eqref{M-vertex} we need to evaluate the diagrams in 
Fig.~\ref{vertex correction} with an operator insertion given by $i\sigma^{\mu\nu}q_{\nu}/2M$, which 
only acts on the $u$ and $d$ quarks. For the kaon the result is
\begin{align} 
&\Lambda^{\mu}_{K,T}(p',p) = 6i\,g_{K}^2 \int \frac{d^4 k}{(2\pi)^4} \no \\
&
\times{\rm Tr} \biggl[\gamma_{5}\,\lambda_{-}\,S_\ell(p'+k)\,\frac{i\sigma^{\mu\nu}q_{\nu}}{2M}\, S_\ell(p+k)\,\gamma_{5}\,\lambda_{+}\,S_s(k)\biggr] \no \\
&= -(p'+p)^{\mu} \frac{6i\,g_{K}^2\,Q^2}{M}\int^{1}_{0}dx \int^{x}_{-x} dy \no \\
&\hs{28mm}
\times \int \frac{d^4 k}{(2\pi)^4} \frac{(1-x)\,M_{s} + x\,M}{[k^2 - \triangle_{4}]^3}.
\label{K-sigma-term}
\end{align}
Setting $M_s = M$ in Eq.~\eqref{K-sigma-term} gives the pion result:
\begin{align}
\Lambda^{\mu}_{\pi,T}(p',p) &= -(p'+p)^{\mu}\, 6i\,g_{\pi}^2\,Q^2 \no \\
&\hs{3mm}
\times \int^{1}_{0}dx \int^{x}_{-x} dy \int \frac{d^4 k}{(2\pi)^4} 
\frac{1}{[k^2 - \triangle_{2}]^3}.
\end{align}
Therefore, the complete result for the pion form factor, including pion cloud effects, is given by
\begin{align}
F_{\pi}(Q^2) &= \lf[Z_{Q}+F_{1Q}^{(q)}(Q^2)+F_{1Q}^{(\pi)}(Q^2)\rg]F_{\pi}^{(\text{bare})}(Q^2) \no \\
&- 6i\,g_{\pi}^2\,Q^2\,\lf[F_{2Q}^{(q)}(Q^2)+F_{2Q}^{(\pi)}(Q^2)\rg] \no \\
&\hs{7mm}
\times \int^{1}_{0}dx \int^{x}_{-x} dy \int \frac{d^4 k}{(2\pi)^4}\ \frac{1}{[k^2-\triangle_{2}]^3}.
\end{align}
The final result for the kaon form factor including the pion cloud effects is:
\begin{align}
&F_{K}(Q^2) = 24i\,g_{K}^2\lf[Z_{Q} +F_{1Q}^{(q)}(Q^2)+F_{1Q}^{(\pi)}(Q^2)\rg] \no \\
&
\times \int\frac{d^4 k}{(2\pi)^4} \int^{1}_{0}dx\, 
\biggl[\frac{-2x}{3[k^2-\triangle_{1}]^2} + \int^{x}_{-x}dy\ \frac{2N_{1}}{3[k^2-\triangle_{4}]^3}\biggr] \no \\
&
- \frac{6i\,g_{K}^2\,Q^2}{M}\lf[F_{2Q}^{(q)}(Q^2)+F_{2Q}^{(\pi)}(Q^2)\rg]\no \\
&\hs{15mm}
\times \int^{1}_{0}dx \int^{x}_{-x} dy \int \frac{d^4 k}{(2\pi)^4} \
\frac{(1-x)\,M_{s} + x\,M}{[k^2 - \triangle_{4}]^3}   \no \\
&
+24i\,g_{K}^2\int\frac{d^4 k}{(2\pi)^4} \int^{1}_{0}dx \no \\
&\hs{18mm}
\biggl[\frac{-x}{3[k^2-\triangle_{3}]^2} + \int^{x}_{-x}dy\ \frac{N_{2}}{3[k^2-\triangle_{5}]^3}\biggr],
\end{align}
where $\triangle_{1}, \ldots, \triangle_{5}$ have been defined in Eqs.~\eqref{triangle1}--\eqref{triangle5}.



\end{document}